\documentclass[prd,superscriptaddress,twocolumn]{revtex4}
\usepackage{amssymb,amsfonts,amsmath}
\usepackage{epsfig,graphicx}\usepackage{mathrsfs}
\usepackage[active]{srcltx}
\usepackage[matrix,frame,arrow]{xypic}

\newtheorem{lemma}{Lemma}
\newtheorem{sublemma}{Sublemma}

 \newtheorem{corollary}{Corollary}

\mathchardef\minus="002D 

\def\Proof{\medskip\par\noindent{\bf Proof. }}
\def\qed{$\,\blacksquare$\par}
\def\<{\langle}
\def\>{\rangle}
 \def\ket#1{| #1 \rangle}

\def\bvec#1{\boldsymbol{\mathrm #1}} 
\def\v#1{{\boldsymbol{\mathrm #1}}}

\def\Brill{\mathsf B} 
\def\Tromb{\mathsf G} 
\def\Ky{\mathsf X} 
\def\Ball{\mathsf{U}}

\def\Sing{\mathsf F}

\def\L2{{\mathcal L}_2}

\def\bk{\bvec{k}}
\def\bbeta{{\bvec{\beta}}}

\def\bx{\bvec x}

\newcommand{\dd}{\operatorname{d}}

\begin{document}
\title{Special Relativity in a Discrete Quantum Universe}
 \author{Alessandro \surname{Bisio}}
\email[]{alessandro.bisio@unipv.it} 
\affiliation{Università degli Studi di Pavia, Dipartimento di Fisica, QUIT Group}
\affiliation{Istituto Nazionale di
  Fisica Teorica e Nucleare, Sezione di Pavia}

\author{Giacomo Mauro \surname{D'Ariano}} \email[]{dariano@unipv.it}
\affiliation{Università degli Studi di Pavia, Dipartimento di Fisica, QUIT Group}
\affiliation{Istituto Nazionale di Fisica Teorica e Nucleare, Sezione
  di Pavia}

\author{Paolo \surname{Perinotti}}
\email[]{paolo.perinotti@unipv.it} 
\affiliation{Università degli Studi di Pavia, Dipartimento di Fisica, QUIT Group}
\affiliation{Istituto Nazionale
  di Fisica Teorica e Nucleare, Sezione di Pavia}

\begin{abstract}
The hypothesis of a discrete fabric of the universe--the ``Planck scale''--is always on stage, since it solves mathematical and conceptual problems in the infinitely small. However, it clashes with special relativity, which is designed for the continuum. Here we show how the clash can be overcome within a discrete quantum theory where the evolution of fields is described by a quantum cellular automaton. The reconciliation is achieved by defining the change of observer as a change of representation of the dynamics, without any reference to space-time. We use the relativity principle, i.e. the invariance of dynamics under change of inertial observer, to identify a  change of inertial frame with a symmetry of the dynamics. We consider the full group of such symmetries, and recover the usual Lorentz group in the relativistic regime of low energies, while at the Planck scale the covariance is nonlinearly distorted.
\end{abstract}
\pacs{11.10.-z,03.70.+k,03.67.Ac,03.67.-a,04.60.Kz}
\maketitle  

\section{Introduction}



 Is the world continuous or discrete? Richard Feynman
  \cite{feynman1982simulating,hey1998feynman} motivated a discrete
  universe as the only way it can be simulated by its own
  constituents, which means by a quantum computer. Einstein himself
  considered a discrete space-time as a possibility, however, he
  complained about the lack of an appropriate mathematical framework
  \cite{stachel1986einstein}. Usually we dismiss discreteness on the
  basis of a mathematical convenience of continuos theories.  But the
  continuum leads to still unsolved mathematical problems in the
  infinitely small, problems that do not arise in the discrete.  The
  discrete, on the other hand, seems to raise a problem: the
  disagreement with Einstein's special relativity.


The debate about the clash between a discrete space and Lorentz
symmetry has been recently renewed because in some approaches to quantum gravity
(such as Regge Calculus \cite{immirzi1997quantum}, spin-foam
\cite{perez2003spin}, causal sets, \cite{bombelli1987space}) the
fundamental description of space-time is a discrete structure, to
which the continuum is only an approximation. The scale of this
discreteness is the Planck length, which is amazingly small--the
Planck length compared to a meter is like the electron radius compared
to the size of our galaxy.

Why the Lorentz transformations would not work with a discrete
space-time? The objection is that a discrete space-time would not be
invariant under the Lorentz group, even if we take it as discrete.
Such a point, raised more than sixty years ago, was disproved for
$d=3$ space dimensions in Ref.  \cite{Schild}. However, it was shown
that the minimum admissible boost would be huge: 0.866 times the speed
of light! It seems therefore that there is no way for the
reconciliation of Einstein's special relativity with a discrete fabric
of space-time. The issue is, however, a false problem, originating
from the unnecessary requirement of enforcing a covariance designed
for the continuum. The right point of view is to take the Lorentz
covariance only as an approximate symmetry, and recovering it in the
regime where the discreteness looks continuous. This is similar to
what happens for a crystalline medium, which looks isotropic at large
scales, whereas instead it is highly anisotropic at microscopic
scales. The smaller is the crystal structure, the more accurate is the
continuum symmetry: think that the Planck length is $10^{-25}$\AA!

The right point of view is thus to consider the continuum as an
approximation of the discrete when observed at very large scales. It
is thus conceptually legitimate that the Lorentz transformations are
actually distorted at the tiny Planck scale. An example of such a
distorted Lorentz symmetry is that of {\em doubly special relativity}
\cite{amelino2001testable,amelino2002relativity,magueijo2002lorentz},
where the distorted Lorentz transformations, in addition to the speed
of light, preserve also an energy scale.

Here we show that if we take only the very essence of the relativity
principle--the invariance of the physical law under change of {\em
  inertial representation}--we get nonlinear Lorentz transformations,
which happen to be of the same kind as those of doubly special
relativity. In the continuum description the reference frame is a
Cartesian coordinate system--what is called ``position
representation'' in quantum theory. Other representations of the
dynamics are given in terms of constants of motion--such as the
momentum or energy representations--and these provide a viable notion
of inertial frame in a world made of countably many quantum systems.

\section{Results}
We take the pragmatic point of view that quantum theory is more efficient a description of our world
than classical theory, pretty much like Kepler laws for the planet orbits are more efficient than
Ptolemaic epicycles. We therefore consider the most general quantum discrete theory, which is a {\em
  quantum cellular automaton}
\cite{feynman1982simulating,schumacher2004reversible,arrighi2011unitarity}.  This consists of
infinitely many quantum systems (qubits, Fermionic or Bosonic fields), whose evolution occurs in
discrete steps, and which interact {\em locally}, namely every system interacts with a bounded 
number of other systems. A consequence of locality is that signals propagate at finite speed over the 
interacting network. For our purposes it is sufficient to consider a single particle, and technically 
this simplifies the automaton to be linear in the quantum field. Moreover we require that the
dynamics is reversible, hence it is described by a unitary matrix: this is what is called {\em
  Quantum Walk} \cite{meyer1996quantum,ambainis2001one}.

In the discrete context universality of the physical law corresponds to the homogeneity of the
evolution. This implies that the network of interactions represents a group $G$ of ``translations'', 
moving from a system to another interacting with it: this is a so-called {\em Cayley graph} of $G$ (see Methods).
Moreover, if we want to reproduce the physical situation
of flat space where parallel transports along two different paths end up in the same result, then
the ``translation'' group must be Abelian. This allows us to make use the Fourier transform. Upon
denoting by $|\bk\>$ the eigenvector of the translations corresponding to wave-vector $\bk$, we
write the quantum walk unitary operator $A$ as follows
\begin{equation}
    A=\int_\Brill\dd\bk\, A_{\bk} \otimes|\bk\>\<\bk|,
\end{equation}
 \begin{figure}[t]
 \includegraphics[width=\columnwidth]{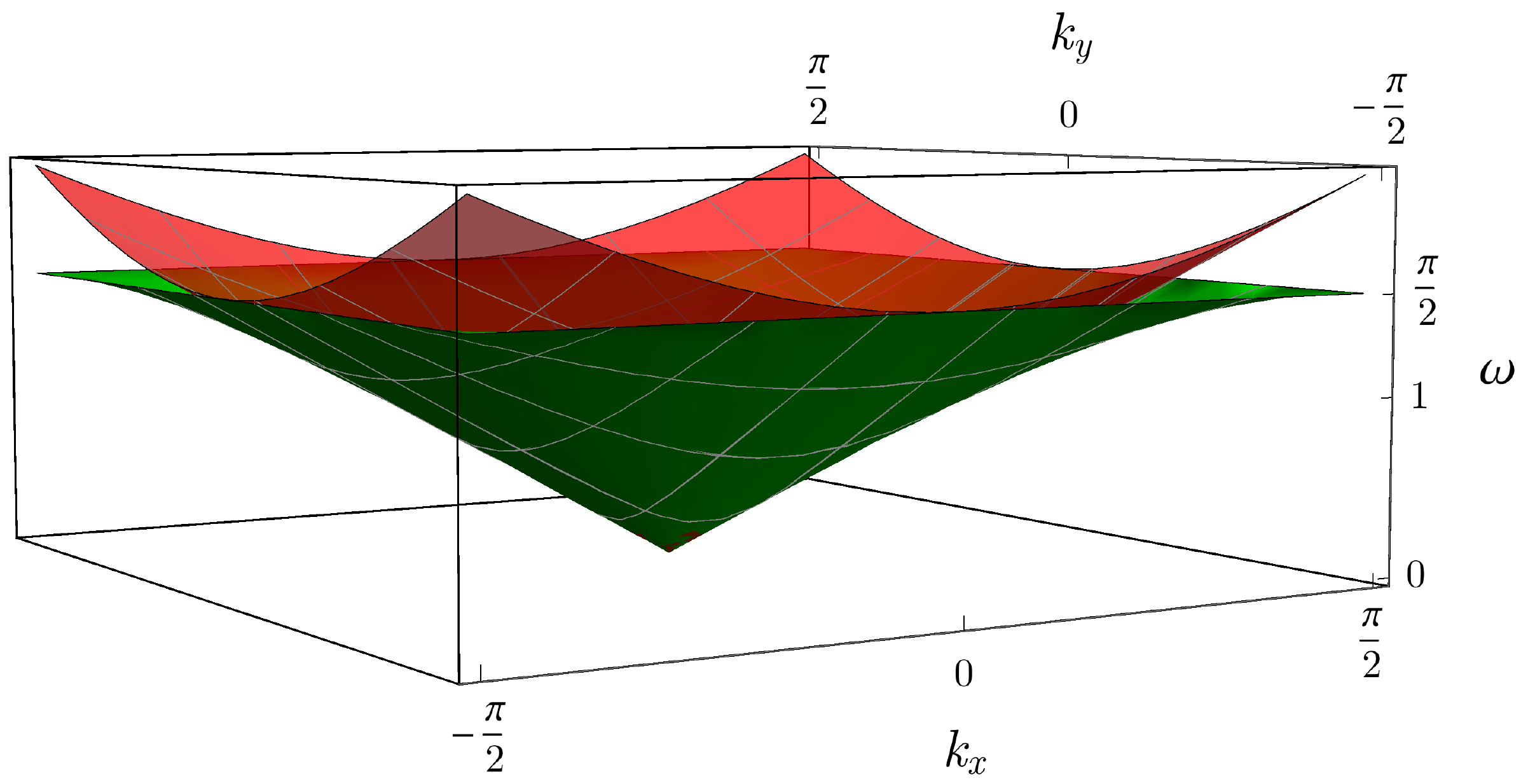}
 \caption{The green surface represents the dispersion relation in Eq. (\ref{eq:eigenequation2}) where we fixed 
 $k_{z}=0$ ($\omega_{\bk}=\arccos\lambda^{\pm}(\bk)$ have the same plot). The red surface is the usual relativistic dispersion 
 relation $\omega_{\bk}^2=k_x^2+k_y^2$. Notice that the two surfaces get closer approaching the origin for $\bk\to 0$.}
   \label{f:disp}
 \end{figure}
where $\Brill$ denotes the Brillouin zone of the interaction lattice, and $A_k$ is a unitary $s$
dimensional matrix, with $s$ finite. As proved in Ref. \cite{PhysRevA.90.062106}, the easiest nontrivial
quantum walk has $s=2$. The eigenvalues of $A$ are obtained by solving the eigenvalue equation
\begin{equation}
A_{\v{k}}\psi(\omega,\v{k})=e^{i\omega}\psi(\omega,\v{k})
\end{equation}
which can be rewritten in relativistic notation as follows
\begin{align}
\label{eq:hamiltonian2}
n_\mu(k)\sigma^\mu\psi(k) = 0,
\end{align}
where we introduced the four-vectors $k=(\omega,\bk)$, $n(k)=(\sin\omega,\v{n}(\bk))$, and 
$\sigma=(I,\v{\sigma})$, with $\v{\sigma}=(\sigma_x,\sigma_y,\sigma_z)$ denoting the Pauli matrices,
and the vector $\v{n}(\bk)$ are defined by
\begin{equation}
\label{eq:generalautomaton}
\v{n}(\v{k})\cdot\v{\sigma}: = \frac{i}{2}(A_{\v{k}} -A_{\v{k}}^\dagger).
\end{equation}
 \begin{figure*}[t]
 \includegraphics[width=\textwidth]{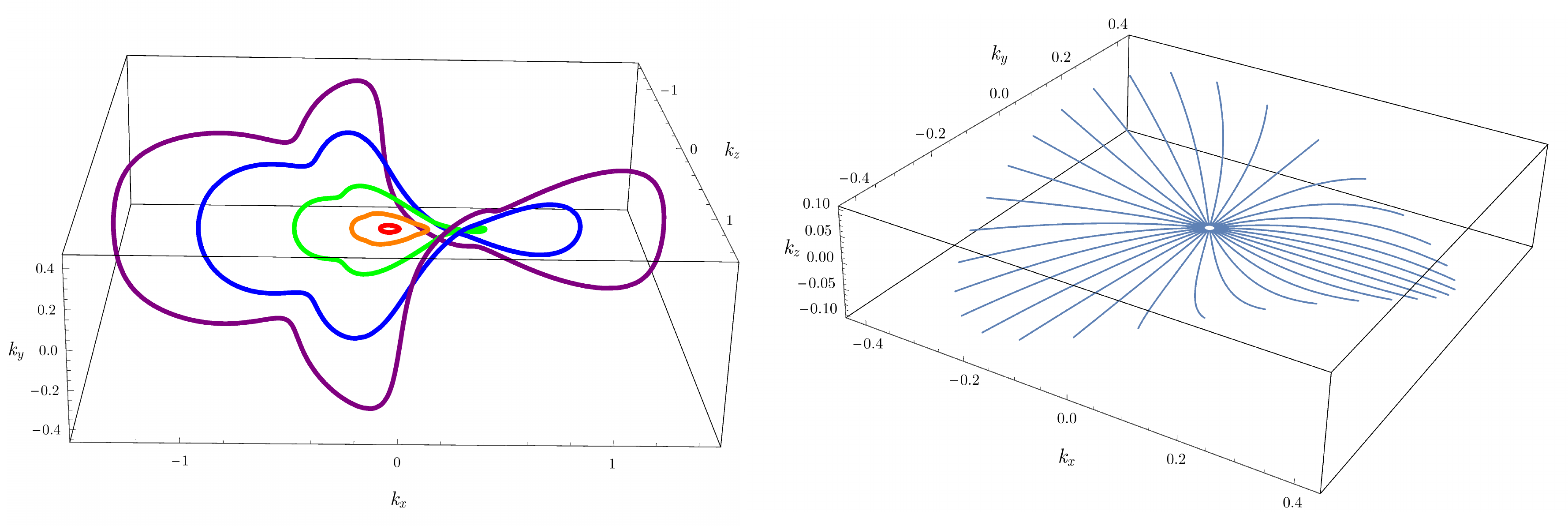}
 \caption{The distortion effects of the Lorentz group for the discrete Planck-scale theory represented by the quantum walk in Eq. (\ref{eq:weyl3D}). 
  Left figure: the orbit of the wavevectors $\bk=(k_x,0,0)$, with $k_x\in\{.05,.2,.5,1,1.7\}$ under the rotation around the $z$ axis. 
 Right figure: the orbit of wavevectors with $|\bk|=0.01$ for various directions in the $(k_x,k_y)$ plane under the boosts with $\bbeta$ parallel to $\bk$ and $|\bbeta|\in[0,\tanh 4]$.}
   \label{orb2d}
 \end{figure*}

The eigenvalues can be collected into two functions $\omega^\pm(\bk)$ called {\em dispersion
  relations}. In this scenario the constants of motions are $\bk$ and $\omega^\pm$, hence a change
of representation corresponds to a map $k\mapsto k'(k)$. Now the principle of relativity corresponds
to the requirement that the eigenvalue equation (\ref{eq:hamiltonian2}) is preserved under a change
of representation as follows
\begin{align}
\label{eq:invariantdynam}
n_\mu(k)\sigma^\mu=
\tilde\Gamma^{-1}_k\,n_\mu(k')\sigma^\mu\,\Gamma_k,
\end{align}
where $\Gamma_k$, $\tilde{\Gamma}_k$ are invertible matrices.

Eq. \eqref{eq:invariantdynam} translates the relativity principle for the QW evolution: the dynamics
is left invariant by a change of observer.

The simplest example of change of observer is the one given by the trivial relabeling $k'=k$ and by
the matrices $ \Gamma_k=\tilde{\Gamma}_k=e^{i\lambda(\v{k})}$, where $\lambda(\v{k})$ is an
arbitrary real function of $\v{k}$.  When $\lambda(\v{k})$ is a linear function we recover the usual
group of translations. The set of changes of representation $k\mapsto k'(k)$ for which Eq.
(\ref{eq:invariantdynam}) holds are a group, which is the largest group of symmetries of the
dynamics. 

If to the general assumptions defining the quantum walk we just add that of isotropy, 
it turns out that there are only two admissible quantum walks \cite{PhysRevA.90.062106}, 
which in the small wave-vector regime give exactly the two Weyl equations for the left and right 
massless Fermion. Indeed, with the above assumptions the only possible lattice is the body
centered cubic one, and modulo local unitary equivalence the two admissible quantum walks are
\begin{equation}\label{eq:weyl3D} 
A^{\pm}_{\bk} := \lambda^{\pm}(\bk) I-i{\bvec{n}}^{\pm}(\bk)\cdot\boldsymbol{\sigma}^\pm,
\end{equation}
where
\begin{align}
&{\bvec{n}}^{\pm}(\bk) :=
\begin{pmatrix}
s_x c_y c_z \pm c_x s_y s_z\\
c_x s_y c_z \mp s_x c_y s_z\\
c_x c_y s_z \pm s_x s_y c_z
\end{pmatrix},
\nonumber\\
&\lambda^{\pm}(\bk) := (c_x c_y c_z \mp s_x s_y s_z ),\;
\label{eq:lambda}\\
&c_\alpha := \cos({k}_\alpha/\sqrt{3}),\;s_\alpha:=
\sin({k}_\alpha/\sqrt{3}),\;\alpha = x,y,z,\nonumber
\end{align}
where $\v\sigma^+=\v\sigma$ and $\v\sigma^-=\v\sigma^T$, with $T$ denoting the transposed
matrix. The dispersion relations are given by 
\begin{align}
\label{eq:eigenequation2}n_\mu^\pm(k)n^{\mu\pm}(k)=0, 
\end{align}
and are plotted in Fig. \ref{f:disp}.

In the small wave-vector regime $\v{k}\sim\v{k}_0=(0,0,0)$ one has $n(k)\sim k$, recovering the
usual relativistic dispersion relation. The Weyl equations can be also recovered in the neighborhood
of the wavevectors $\v{k_1}=\frac{\pi}{2}(1,1,1)$, $\v{k_2}=-\frac{\pi}{2}(1,1,1)$,
$\v{k_3}=-\frac{\pi}{2}(1,0,0)$. The mapping between the vectors $\v{k}_i$ exchange chirality of the
particle and double the particles to four species in total. Therefore we have four different
particles--two left-handed and two right-handed--namely the discreteness also doubles the particles,
which is the well known phenomenon of Fermion doubling \cite{PhysRevD.16.3031}. In the following the term ``small
wavevector'' will denote the neighborhoods of the vectors $\v{k}_i$ $i=0,\ldots3$.

\begin{figure}[t]
 \includegraphics[width=\columnwidth]{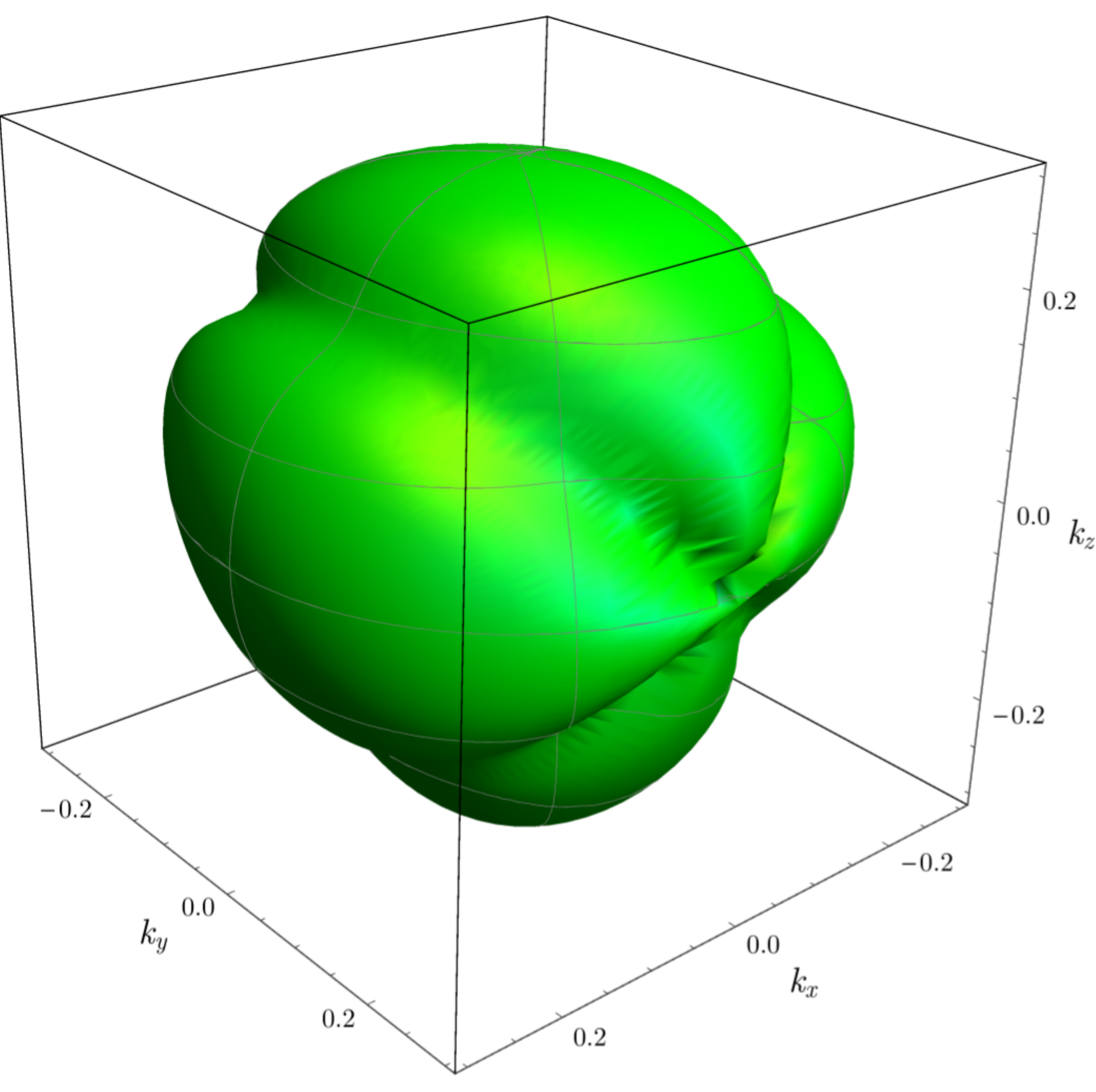}
 \caption{The green surface represents the orbit of the wavevector $\bk=(0.3,0,0)$ under the full rotation group $SO(3)$.}
   \label{orb3d}
 \end{figure}

We now show that the group of symmetries of the dynamics of the quantum walks in Eq.
(\ref{eq:weyl3D}) contains a nonlinear representation of the Poincar\'e group, which exactly
recovers the usual linear one in the small wave-vector regime.  For any arbitrary non vanishing
function $f(k)$ we can introduce the four-vector
\begin{align}
\label{eq:fff}
p^{(f)}=\mathcal{D}^{(f)}(k):=f(k)n(k) 
\end{align}
and rewrite the eigenvalue equation (\ref{eq:hamiltonian2}) as follows 
\begin{align}
  p^{(f)}_\mu \sigma^{\mu} \psi(k) = 0.
\end{align}
Upon denoting the usual Lorentz transformation by $L_\bbeta$ for a suitable $f$ (an example is
provided in the supplemental material) the Brillouin zone splits into four regions $\Brill_i$
$i=1,\ldots,4$ centered around $\v{k}_i$ $i=0,\ldots 3$, such that the composition
\begin{align}\label{calL}
  \mathcal{L}^{(f)}_\bbeta := \mathcal{D}^{(f)-1} L_\bbeta \mathcal{D}^{(f)}
\end{align}
is well defined on each region separately (see Methods). The four invariant regions corresponding to the four
different massless Fermionic particles show that the Wigner notion of ''particle'' as invariant of
the Poincar\'e group survives in a discrete world, consistent with a physical interpretation of the
Fermion-doubled particles. For fixed function $f$ the maps $\mathcal{L}^{(f)}_\bbeta$ provide a
non-linear representation of the Lorentz group
\cite{amelino2001planck,amelino2002relativity,magueijo2002lorentz}. 
In Figs. \ref{orb2d} and \ref{orb3d} we show the numerical evaluation of some wavevector orbits 
under subgroups of the nonlinear Lorentz. The distortion effects due to underlying discreteness 
are evident at large wavevectors and boosts. The relabeling $k\rightarrow k'(k)=\mathcal{L}^{(f)}_\bbeta(k)$ satisfies (\ref{eq:invariantdynam}) with $\Gamma_k=\Lambda_\bbeta$
and $\tilde\Gamma_k=\tilde\Lambda_\bbeta$ for the right-handed particles, and
$\Gamma_k=\tilde\Lambda_\bbeta$ and $\tilde\Gamma_k=\Lambda_\bbeta$ for the left-handed particles,
with $\Lambda_\bbeta$ and $\tilde\Lambda_\bbeta$ being the $(0,\tfrac12)$ and $(\tfrac12,0)$
representation of the Lorentz group, independently on $k$ in each
pertaining region.

For varying $f$, we obtain a much larger group, including infinitely many copies of the nonlinear
Lorentz one. In the small wave-vector regime the whole group collapses to the usual linear Lorentz
group for each particle. 

Up to now we have analyzed what happens with massless particles. A simple way to obtain the Dirac
equation is to pair an automaton in Eq. (\ref{eq:weyl3D}) with its adjoint
into a direct sum, as in Ref.\cite{PhysRevA.90.062106},
 thus leading to a new automaton giving the Dirac equation in the small wave-vector regime.
A relevant feature of the discreteness is that because of unitarity the mass parameter is upper
bounded \cite{mauro2012quantum}.  Now if one derives the full symmetry group of the dynamics as we have done for
the two automata in Eq. (\ref{eq:weyl3D}) one discovers that the group is a nonlinear representation of
the de Sitter group $SO(1,4)$ with infinite cosmological constant, with the rest mass of the
particle playing the role of the additional coordinate (see Methods). It is noticeable
that even for pure boosts the rest-mass is involved in the transformation. For rest-mass much
smaller than the upper bound and for pure boosts one recovers the previous nonlinear Lorentz group
for zero-mass.

\section{Conclusion}
We have seen what happens of the Lorentz group in a quantum world that is discrete. The main point
is to abandon the idea of enforcing the exact Lorentz symmetry on the discrete, but instead to
consider the symmetry as an approximate one that holds only in the small wave-vector and small mass
regime. But the natural question is now: how small? According to the common opinion the scale of
discreteness $a$ is identified with the Planck scale. In terms of the maximum wavevector $k_M$ in
the Brillouin zone, one has $k_M=\frac{\sqrt3\pi}{a}$. In the small wavevector regime we recover
the simple relations \cite{mauro2012quantum} $c=\frac{a}{\sqrt3 \tau}$ and $\hbar=\mu a c$, with $c$, $\hbar$,
and $\tau$ denoting the speed of light, the Planck constant, and the time-step, respectively. Then
the maximum mass $\mu$ of the quantum walk is the Planck-mass.  A way of deriving $\mu$ and $a$
heuristically is to keep literally the argument of taking the mass of the particle bounded in
order to keep the Compton wave-length $\lambda_C$ larger than the Schwartschild radius.  Noticeably
for $m=\mu$ the dispersion relation is constant, namely with no propagation of information, a
situation reminiscent of a micro black hole \cite{PhysRevD.53.3099}. Remarkably general relativity
enters the present quantum digital framework also through the unforeseeable appearance of the De
Sitter symmetry group, which connects different Dirac particle mass values. Are these only
coincidences? The dream is that it is a new route to quantum gravity. 

The crucial question is now what can be actually seen experimentally. 
Recently, experimental tests of Planck-scale phenomenology
have been proposed \cite{Moyer:2012ws,Hogan:2012ik,pikovski2011probing}.
In particular, the modification to the usual
dispersion relations can in principle be detected in observation of gamma-ray bursts from deep-space
events, where billions of light-year of distance can sufficiently amplify the weak vacuum dispersive
behavior due to discreteness \cite{Amelino-Camelia:1998aa}. In our context this can be proved with the free
electromagnetic field derived as the two-particle sector of the quantum walk in Eq. (\ref{eq:weyl3D}) \cite{Bisio2016}.
This possibility reconnects with the recent analysis of data \cite{Vasileiou:2015} from Fermi-LAT concluding that 
the observations set an upper-bound to the scale of discreteness which is smaller than the Planck scale $a$ by a factor
$2.8$. The analysis of Ref.  \cite{Vasileiou:2015} can be refined with a complete theoretical derivation based on Ref. 
\cite{Bisio2016} and on the results presented in this letter. This would also take into account the possibility of 
a compensating effect due to the phenomenon of relative locality \cite{PhysRevD.84.084010}. 
In short relative locality is the phenomenon due to the 
nonlinearity of the Lorentz transformations, which generalizes the relativity of simultaneity to relativity 
of the full space-time coincidence of events. The separation of events under boost is amplified by the 
difference of their frequency domain. Indeed the Fermi-LAT observation is based on a predicted time-delay 
between two events with a huge difference in frequency, which could then be compensated by the 
relative-locality effect. The fully fledged discrete theory given here, derived from very general principles, 
allows for a thorough quantitative evaluation that takes into account both the dispersive vacuum and nonlinear 
Lorentz transformations.

\acknowledgments This work has been supported by the Templeton Foundation under the project ID\# 43796 {\em A Quantum-Digital Universe}.

%

\clearpage
\onecolumngrid
\begin{center}
\begin{large}
\textbf{SUPPLEMENTAL INFORMATION}
\end{large}
\vspace{0.7cm}
\end{center}
\twocolumngrid


\section{Quantum Walk from principles}
Quantum Walks (QWs) describe the evolution of a quantum particle over
a lattice. The dynamics is assumed to be \emph{reversible}, hence the
QW will be represented by a unitary transformation.  By denoting with
$G$ the set of the lattice's points we can convenienty introduce the
Hilbert space $\ell^{2}(G)$ and the orthonormal basis $\ket{\bvec{x}}$
which corresponds to the position of the particle.  If we associate
the Hilbert space $\mathbb{C}^s$ to the internal degrees of freedom of
the particle, the QW is then a unitary operator on
$\ell^{2}(G) \otimes \mathbb{C}^s$.

\paragraph{Locality}
The QW evolution is assumed to be \emph{local} i.e. information
propagates through the lattice at a bounded speed.  Given a lattice,
let $N_{\bvec{x}}$ be the set of nearest neighbors of the site
$\bvec{x}$. If the particle is localized at site $\bvec{x}$, then,
after one step of the QW, it must be localized within a finite set
$N_{\bvec{x}}$.  Such a locality condition introduces a notion of
causal cone in the lattice.
\paragraph{Homogeneity}
The evolution is assumed to be \emph{homogeneous}, i.e.  requirement
that all the sites are equivalent. The QW evolution should not not
allow one to discriminate between two points $\bvec{x}$ and
$\bvec{x}'$. This requirement (see Ref. \cite{dirac3d} for
a full derivation and Appendix \ref{sec:weyl-automaton-3d} for a shorter review)
implies that the set og points $G$ is a group and the lattice is a
possible Cayley Graph of this group.
\paragraph{Isotropy}
The assumption of \emph{isotropy} translate the requirement that there
is no priviledged direction on the lattice. The mathematical
traslation of this requirement requires the existence of a group of
permutation that act on the generators of the group $G$ that can be
faithfully represented on the internal degrees of freedom (see
Ref. \cite{dirac3d} for a full derivation and the
Appendix \ref{sec:weyl-automaton-3d} for a shorter review)

\paragraph{Flat and curved space}
The above sketched framework encompass a broad variety of dynamics. In
particular, depending on the properties of the group $G$, we can have
a quantum dynamics on a generally curved space. 
If we are interested to make contact with special relativity, it is
natural to restrict the scenario to QWs corresponding to dynamics on
the flat Minkovski spacetime. This requirement correspond to assume
the group $G$ to be \emph{virtually abelian}, i.e. $G$ has an abelian
subgroup of finite order. We can further restrict ourselves to the case in which $G$
is \emph{abelian} without any loss of generality. The price to pay for
this restriction is to add additional internal degrees of freedom (see
Ref. \cite{d2015virtually2} and Appendix \ref{sec:weyl-automaton-3d} for a more
complete discussion).

\paragraph{Fourier analysis}
If the group $G$ is abelian it is convenient to study the dynamics in
the Fourier transform basis $| \bvec k \> := (2
\pi)^{\frac{3}{2}}\sum_{\v{x}} e^{i \v{k}\cdot \bx} \ket{\bx}$.
Since homogeneity condition implies that the QW commutes with
the translations  on the lattice, in the Fourier basis the QW operator
can be written as
\begin{align}
  \label{eq:Sweylautomata} 
  A = \int_B\operatorname d^3 \! \bk  \,  |{\bk}\>\< {\bk}| \otimes
  A_{\bk}
\end{align}
where $B$ denotes the first Brillouin zone of the underlying lattice.
The unitary constraint implies that 
$A_{{\bk}}$ is unitary for every ${\bk}\in B$ and
the locality assumption implies that 
$A_{{\bk}}$ is a matrix polynomial in $e^{i \bvec{h} \cdot \bvec{k}}$.
Notice that due to the discreteness of the lattice the QW is
band-limited in ${\bk}$.
The spectrum $\{e^{i\omega^{(i)}_{\bk}}\}$ of the operator $ A_{\bk}$, and
expecially its {\em dispersion relation} that is the
expression of the phases $\omega^{(i)}_{\bk}$ as functions of ${\bk}$,
plays a crucial role in the analysis of the QW dynamics. 
\paragraph{Weyl Quantum Walk}
If the dimension of the Hilbert space of the internal degrees
of freedom is $s=2$ and  the group $G$ is $\mathbb{Z}^3$
the requirements  of \emph{locality}, \emph{homogeneity} and
\emph{isotropy} then \cite{dirac3d} the
 QWs can only be defined over the
body-centered-cubic lattice and they are equivalent (up to a local
change of basis) to the following two QWs:

\begin{align}\label{eq:Sweyl3Dbis} 
A^{\pm}_{{\bk}} &:= \lambda^{\pm}({\bk}) I-i{\bvec{n}}^{\pm}({\bk})\cdot\boldsymbol{\sigma}^\pm,\\
{\bvec{n}}^{\pm}({\bk}) &:=
\begin{pmatrix}
s_x c_y c_z \pm c_x s_y s_z\\
c_x s_y c_z \mp s_x c_y s_z\\
c_x c_y s_z \pm s_x s_y c_z
\end{pmatrix},
\nonumber\\
\lambda^{\pm}({\bk}) &:= (c_x c_y c_z \mp s_x s_y s_z ),\;
\label{eq:Slambdabis}\\
c_\alpha &:= \cos({k}_\alpha/\sqrt{3}),\;s_\alpha:=
\sin({k}_\alpha/\sqrt{3}),\;\alpha = x,y,z.\nonumber
\end{align}
The Pauli matrices $\v\sigma^+=\v\sigma$ are the usual ones, while the
$\v\sigma^-=\v\sigma^T$ are their transposed ones, 
and ${\bk} \in \Brill$ where $\Brill$ denotes the Brillouin zone of the
BCC lattice.

\section{Nonlinear Lorentz transformations for the Weyl quantum
  walk}

In this section we give the explicit construction of the non-linear representation
of the Lorentz group on the set of the solutions of the Weyl QW
dynamics.  
The set of solution will be split into four regions, each one of them
carrying a non-linear deformation of the Lorentz group.

Let us consider the splitting of the Brillouin zone $\Brill$
\begin{align}
    \label{eq:Sprezones}
  \begin{aligned}
    \Brill'_0 &:=  \{\v{k}\in\Brill |\lambda(\v{k}) > 0, \cos(2 k_y/\sqrt{3}) > 0\},\\
 \Brill'_1 &:=  \{\v{k}\in\Brill|\lambda(\v{k}) < 0, \cos(2 k_y/\sqrt{3}) > 0\},\\
 \Brill'_2 &:=  \{\v{k}\in\Brill|\lambda(\v{k}) > 0, \cos(2 k_y/\sqrt{3}) < 0\},\\
 \Brill'_3 &:=  \{\v{k}\in\Brill|\lambda(\v{k}) < 0, \cos(2 k_y/\sqrt{3}) < 0 \},
  \end{aligned}
\end{align}
with $\Brill= \cup_{i=0}^{3} \Brill'_i$ up to a nullmeasure set,
and let us denote with $\v{n}^{(i)}(\v{k})$
the restriction of $\v{n}(\v{k})$ to $\Brill'_i$. Notice that we dropped the label
$\pm$ denoting the two different Weyl walk, since the results holds
the same in the two cases. 
We now denote with $\mathsf{U}$ the unit open ball in $\mathbb{R}^3$
($\overline{\mathsf{U}}$ denotes its closure) and with $\mathsf{H}$
the subset of  $\mathsf{U}$ defined as follows
\begin{align}
  \mathsf{H} := \{ \v{m}\in \mathsf{U} s.t. m_x = \pm m_z , 2m_x^2 +
  2m_y^2 \geq 1   \}.
\end{align}
We then consider the regions
\begin{align}
    \label{eq:Szones}
  \begin{aligned}
    \Brill_i & := \v{n}^{(i)-1}(\mathsf{U} \setminus  \mathsf{H})
  \end{aligned}
\end{align}
and the function $f(\omega, {\bk}) $
defined as follows:
\begin{align}
\begin{aligned} 
f(\omega, {\bk}) &=  g(\v{n}({\bk})) = \\
=&\tilde{g}(r,\theta,\phi) := 1+ r \!\!\int_0 ^r \!\!\!\! ds \,\,\left(
  \frac{1}{a(s)} + \frac{1}{b(s,
    \theta, \phi)} \right)\\
a(r) &:=1- r^2 , \\
 b(r,\theta, \phi) &:= (\cos^2(\phi) - \sin^2(\phi))^2 + \\
&+ (\tfrac{1}{2} - r^2 (1- \cos^2(\theta) \sin^2(\phi)))^2
  \end{aligned} 
  \end{align}
where we used the spherical coordinates
$n_x(\v{k}) = r \cos\theta \cos\phi$,
$n_y(\v{k}) = r \sin\theta $,
$n_z(\v{k}) = r \cos\theta \sin\phi$.
Finally, we define the maps $\mathcal{D}^{(i)}$ as
\begin{align}
\label{eq:Stheultimatedeformation2}
\begin{aligned}
& \mathcal{D}^{(i)} :  \mathsf{\Sigma}_i \to \mathsf{\Gamma}_0, \quad
  \mathcal{D}^{(i)} : 
\begin{pmatrix}
 \omega\\
\v{k}
\end{pmatrix}
 \mapsto 
f(\omega, {\bk})
\begin{pmatrix}
  \sin\omega\\
\v{n}^{(i)}(\v{k})
\end{pmatrix},\\
&\mathsf{\Sigma}_i := \{ (\omega , \v{k}) \text{ s.t. }   \v{k} \in
\Brill_i , \sin^2\omega - |\v{k}|^2 = 0 \},\\
&\mathsf{\Gamma}_0 := \{ p\in \mathbb{R}^4 \mbox{ s.t. }  p_\mu p^\mu
=0\}.
\end{aligned}
\end{align}
One can prove (see Appendix ) that the maps  $\mathcal{D}^{(i)}$
define an analytic diffeomorphism between 
 $ \mathsf{\Sigma}_i$ and $ \mathsf{\Gamma}_0 $.
As a consequence, the composition 
\begin{align}
  \label{eq:Stheultimatelorentz}
  \mathcal{L}^{(i)}_{\beta}: \mathsf{\Sigma}_i \to \mathsf{\Sigma}_i \qquad 
 \mathcal{L}^{(i)}_{\beta}  := \mathcal{D}^{-1} \circ L_\beta  \circ  \mathcal{D}
\end{align}
is  a well defined nonlinear representation of the Lorentz group on
the set  $\mathsf{\Sigma}_i$.
Since the union of the $\Brill_i$ sets
coincides with the whole (up to a null measure set) Brillouin zone, we
have that the collection of the maps
  $ \mathcal{L}^{(i)}_{\beta}$ provide a notion of Lorentz
  transformation for any (up to a null measure set) solution of the
  Weyl QW dynamics.

\section{Dynamical symmetries of Dirac Quantum Walk}

The Dirac quantum walk is obtained by performing the direct-sum 
of two Weyl walks and introducing off-diagonal blocks in such a way
that the obtained matrix is a well defined quantum walk. As proved in
Ref. \cite{dirac3d},
 there are
only two admissible walk
\begin{align}
  \begin{aligned}
   &  D^{\pm} = 
  \begin{pmatrix}
    n A^{\pm} &im I \\
    imI & n A^{\pm}  \\
  \end{pmatrix} \\
&0\leq n,m \leq 1 \quad n^2+m^2 =1
  \end{aligned}
\end{align}
The eigenvalue equation of the Dirac QW can be written as 
\begin{equation}\label{eq:Sdesitcov}
[p^{(f)}_\mu(\omega,\v{k},m)\gamma^\mu-mI]\psi(\omega,\v{k},m)=0,
\end{equation}
where $\gamma^{\mu}$ are the Dirac $\gamma$ matrices in the Weyl
representation, and $m$ is then interpreted as the particle mass.  Due
to the explicit dependence of $p^{(f)}_{\mu}$ from $m$ the covariance
under change of reference cannot leave the value of $m$ invariant. In
such case the dispersion relation resorts to the conservation of the
de Sitter norm
\begin{equation}\label{eq:Sdesit}
\sin^2\omega-(1-m^2)|\v{n}(\v{k})|^2-m^2=0. 
\end{equation}
The group leaving Eq. (\ref{eq:Sdesit}) invariant is the De Sitter group $SO(1,4)$. In the limit of $m\ll 1$,
the usual Lorentz symmetry is recovered. The analysis of De Sitter covariance of Eq.
(\ref{eq:Sdesitcov}) will be given in a forthcoming publication.

\appendix

\section{3-dimensional Quantum Walks and the Weyl Quantum Walk}
\label{sec:weyl-automaton-3d}

We consider Quantum Walks (QWs) that describe the evolution of a quantum particle over a
lattice of $\mathbb{R}^3$. The dynamics is assumed to be
reversible, hence the QW will be represented by a unitary transformation.
We consider \emph{Bravais
  lattices}, i.e. lattices that are generated by linear
independent discrete translations.  By
denoting with $G$ the set of the lattice's points we can convenienty
introduce the Hilbert space $\ell^{2}(G)$ and the orthonormal basis
$\ket{\bvec{x}}$ which corresponds to the position of the particle.
 The internal degrees of freedom of the particle are
described by the Hilbert space $\mathbb{C}^s$.

The QW evolution is assumed to be \emph{local} i.e. information
propagates through the lattice at a bounded speed. 
Given a Bravais lattice, let $N_{\bvec{x}}$ be the set of nearest neighbors of the site 
 $\bvec{x}$. If the particle is
localized at site $\bvec{x}$, then, after one step of the QW, it must
be localized within the set $N_{\bvec{x}}$.
The set $N_{\bvec{x}}$  corresponds
to set of vectors $S := \{\bvec{h}\}$ such that
$N_{\bvec{x}} = \{\bvec{y} \mbox{ s.t. }
\bvec{y}=\bvec{x}+\bvec{h},\bvec{h}\in S \}$
Such a locality condition introduces a notion of causal cone in the
lattice.

The QW evolution is also assumed to be
\emph{homogeneous}, i.e. it must commute with the discrete
traslations.
 Let $T_{\bvec{h}}$ be the translation operators on $\ell^{2}(G)$,
such that  i.e.
$T_{\bvec{h}} \ket{\v{x}} =  \ket{\v{x}+\v{h}} $
These assumption implies that the QW evolution can be represented by
a unitary operator on $\ell^{2}(G) \otimes \mathbb{C}^s$ as follows:
\begin{align}
\label{eq:Sgeneralqw}
A := \sum_{\bvec{h}\in S} T_{\bvec{h}}  \otimes A_{\v{h}} 
\end{align}
where $ A_{\v{h}}$ are $ s \times s $ matrices
and $T_{\bvec{h}}$ is the translations that connect the site
$\bvec{x}$ with its nearest neighbor $\bvec{x}+\bvec{h}$.

The QW dynamics is conveniently study in the Fourier transform basis
$| \bvec k \> := (2 \pi)^{\frac{3}{2}}\sum_{\v{x}} e^{i \v{k}\cdot \bx} \ket{\bx}$,
in which the QW operator
$A$  becomes
\begin{align}
  \label{eq:Sweylautomata} 
  A = \int_B\operatorname d^3 \! {\bk}  \,  |{{\bk}}\>\< {{\bk}}| \otimes
  A_{{\bk}},\qquad A_{{\bk}}:=\sum_{\bvec{h} \in S}\,e^{-i{\bk}\cdot\bvec{h}}A_{\bvec{h}},
\end{align}
where
 $B$ denotes the first Brillouin zone of the underlying lattice and it is
defined  
by the following linear constraints:
\begin{equation}
  \begin{split}
&B:=\bigcap_{i }\{{\bk}\in\mathbb R^3|-\pi|\tilde
{\bvec{h}}_i|^2\leq
|{\bk}\cdot\tilde{\bvec{h}}_i|\leq\pi|\tilde{\bvec{y}}_i|^2\}\\
&\tilde{\bvec{h}}_i\cdot\bvec{h}_j=\delta_{ij},\quad  \bvec{h}_j \in S.
  \end{split}
  \label{eq:Sbrill}
\end{equation} 
The unitary constraint implies that 
$A_{\bk}$ is unitary for every ${\bk}\in B$ and
the locality assumption implies that 
$A_{\bk}$ is a matrix polynomial in $e^{i \bvec{h} \cdot \bvec{k}}$.
Notice that due to the discreteness of the lattice the QW is
band-limited in ${\bk}$.
The spectrum $\{e^{i\omega^{(i)}_{\bk}}\}$ of the operator $ A_{\bk}$, and
expecially its {\em dispersion relation} that is the
expression of the phases $\omega^{(i)}_{\bk}$ as functions of ${\bk}$,
plays a crucial role in the analysis of the QW dynamics.

\begin{figure}[t]
 \includegraphics[width=.35\textwidth]{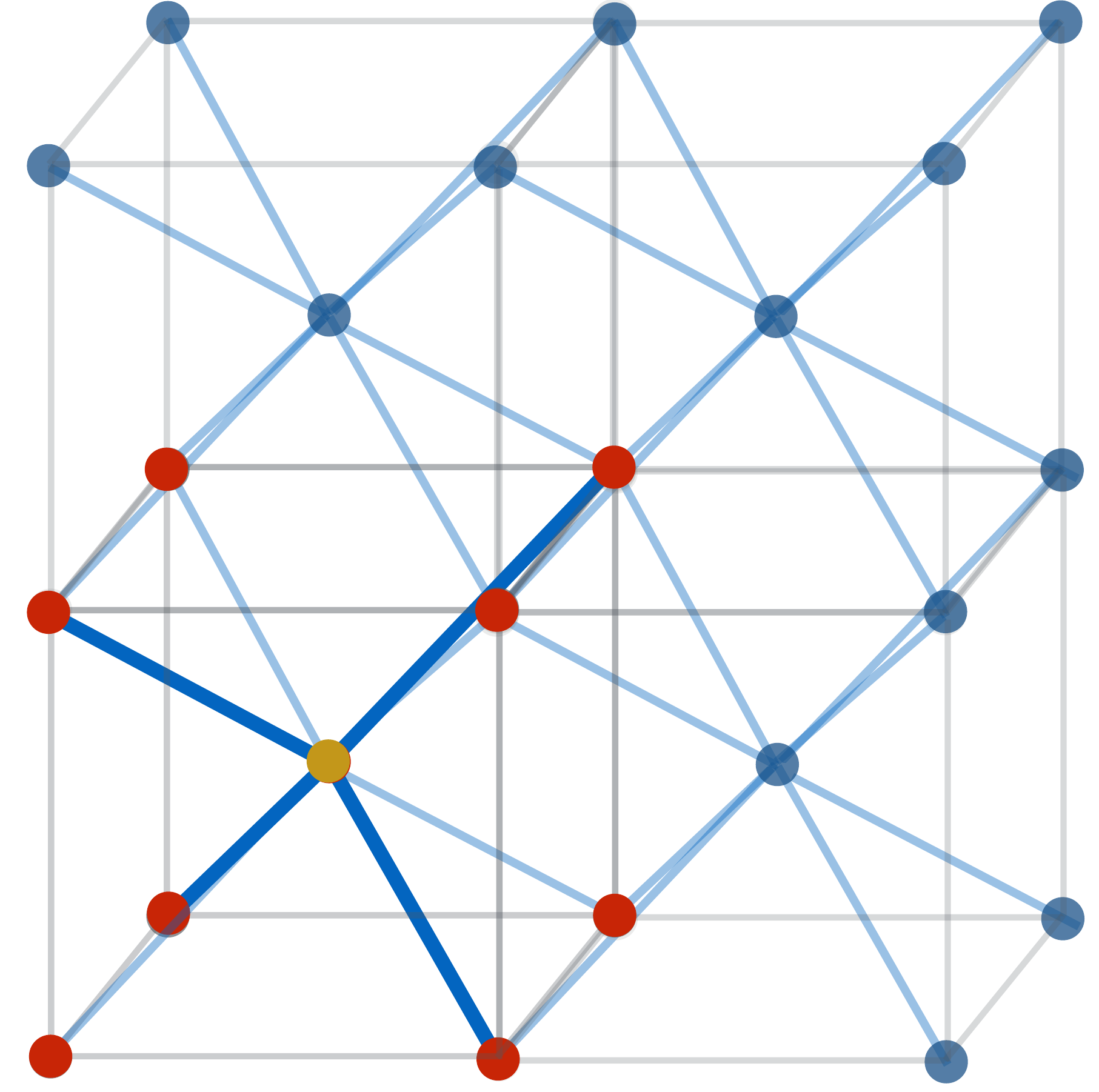}
 \includegraphics[width=.4\textwidth]{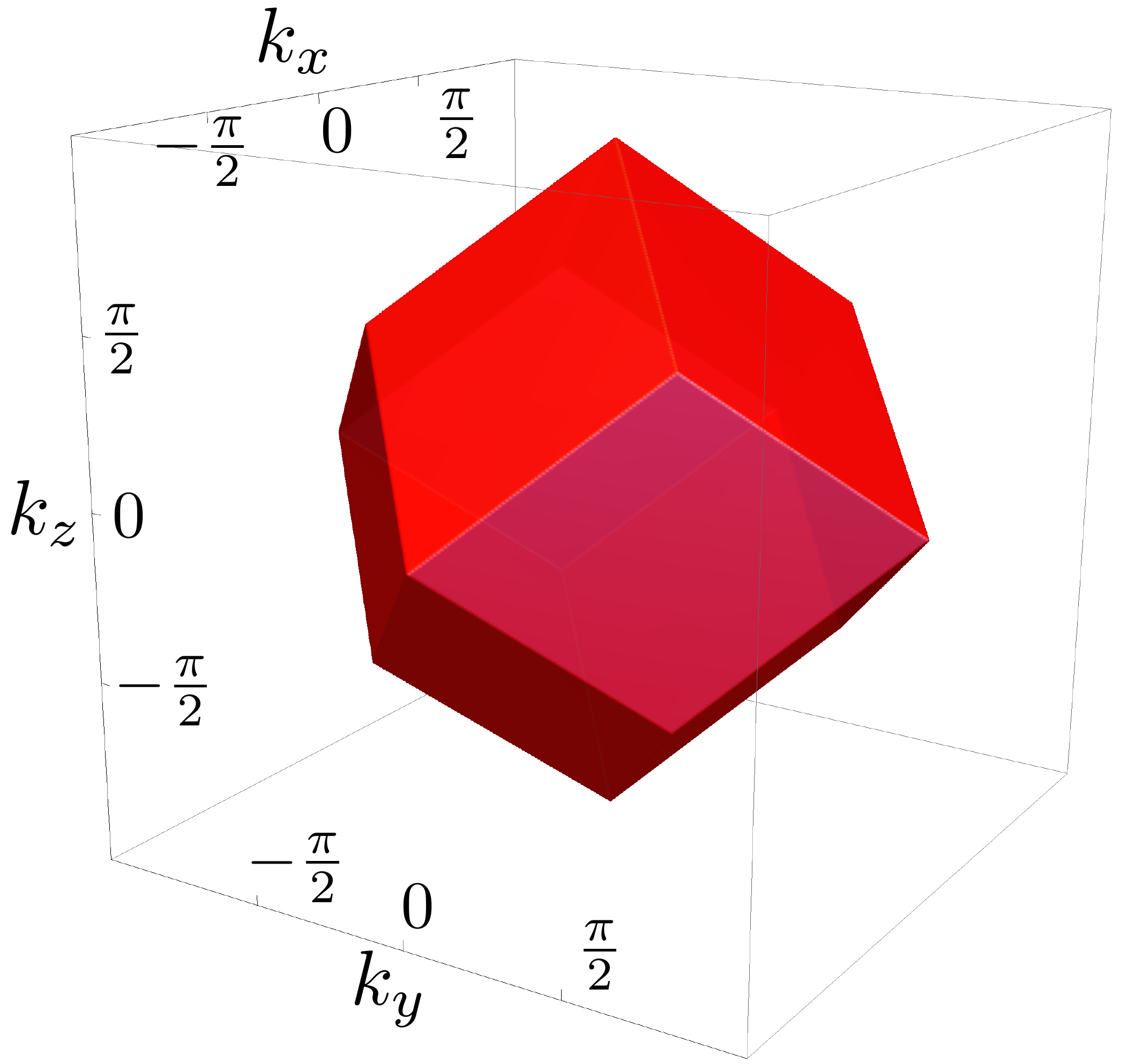}
 \caption{\emph{Top}: Body centered cubic lattice. Each pair of
   nearest neighbor is connected by a blue link. Each point of the
   lattice has $8 $ 
   nearest neighbors, e.g. the red points are the nearest neighbors of
   the yellow one.  
\emph{Bottom}: Brillouin zone $\Brill$ of the BCC lattice. The zone is a rhombic dodecahedron in which the opposite faces are identified.
 }
   \label{Brillouinzone}
 \end{figure}

It was shown in Ref.\cite{dirac3d}
that, when the dimension of the Hilbert space of the internal degrees
of freedom is $s=2$, the additional assumption of \emph{isotropy}\cite{note1}
of the evolution implies that the QWs can only be defined over the
body-centered-cubic lattice and they are equivalent (up to a local
change of basis) to the following two QWs:

\begin{equation}\label{eq:Sweyl3Dbis} 
A^{\pm}_{{\bk}} := \lambda^{\pm}({\bk}) I-i{\bvec{n}}^{\pm}({\bk})\cdot\boldsymbol{\sigma}^\pm,
\end{equation}
where we define
\begin{align}
&{\bvec{n}}^{\pm}({\bk}) :=
\begin{pmatrix}
s_x c_y c_z \pm c_x s_y s_z\\
c_x s_y c_z \mp s_x c_y s_z\\
c_x c_y s_z \pm s_x s_y c_z
\end{pmatrix},
\nonumber\\
&\lambda^{\pm}({\bk}) := (c_x c_y c_z \mp s_x s_y s_z ),\;
\label{eq:Slambdabis}\\
&c_\alpha := \cos({k}_\alpha/\sqrt{3}),\;s_\alpha:=
\sin({k}_\alpha/\sqrt{3}),\;\alpha = x,y,z.\nonumber
\end{align}
The Pauli matrices $\v\sigma^+=\v\sigma$ are the usual ones, while the
$\v\sigma^-=\v\sigma^T$ are their transposed ones, 
and ${\bk} \in \Brill$ where $\Brill$ denotes the Brillouin zone of the BCC lattice 
(see Fig.~\ref{Brillouinzone}).

In the limit $|{{\bk}}|\ll 1$
  we have
\begin{equation}
\bvec{n}^{\pm}({{\bk}})\sim\tfrac{{{\bk}}}{\sqrt{3}},\quad A^{\pm}_{{\bk}}\sim\exp[-i\tfrac{{{\bk}}}{\sqrt{3}} \cdot\boldsymbol{\sigma}^\pm],
\end{equation}
corresponding to the evolution of the Weyl's equation with the
rescaling  $\tfrac{{{\bk}}}{\sqrt{3}} \to {\bk}$.
Since the $A^{+}$ and $A^{-}$ reproduce the dynamics of the
Weyl equation in the  limit $|{{\bk}}|\ll 1$, we refer to them as
\emph{Weyl QW}. 

For the sake of simplicity, in the following we will consider only one
Weyl QW, i.e.~we define $A_{{\bk}} := A_{{\bk}}^+$, the choice of
Pauli matrices is the usual one $\v\sigma:=\v\sigma^+$, and we
similarly drop all the $\pm$ superscripts. This choice is completely
painless since all the results of this paper can be easily adapted to
the choice $A_{{\bk}} = A_{{\bk}}^-$. In order to simplify the notation we
also adopt the convention
\begin{align}
  k_i \to \frac{k_i}{\sqrt{3}}
\end{align}
in order to get rid of the annoying $\frac{1}{\sqrt{3}}$ factor.
With this choice we have that in the $| {\bk} | \ll 1$  limit,
$\v{n}({\bk}) \simeq {\bk}$.

\section{Deformed relativity and Weyl quantum walk}
\label{sec:deform-relat-weyl}

In this section we prove that the construction in Eq. 
is a well defined deformed Lorentz symmetry for each set $\mathsf{\Sigma}_i$
Let us define
\begin{align}
\label{eq:SdefPN}
  & \mathcal{D} = \mathcal{N} \circ \mathcal{P}, \qquad
 \mathcal{P} :
(\omega, \v{k}) \mapsto (\omega,\v{n}(\v{k}) ) \\
& \mathcal{N} :
\begin{pmatrix}
    \omega \\
\v{m}
  \end{pmatrix}
\mapsto
g(\omega,\v{m})
\begin{pmatrix}
    \sin\omega \\
    \v{m}
  \end{pmatrix}
\nonumber
\end{align}
where we also assumed $f(\omega,\v{k}) = g(\omega,\v{n}(\v{k})) $.

We now need to study separately the properties of the two maps
$\mathcal{P}$ and $\mathcal{N}$.

\subsection{Study of the map $\v{n}(\v{k})$}

\label{sec:study-mathcalp}
In this subsection we study the analytical properties of the map
$\mathcal{P}$, which, according to Eq. (\ref{eq:SdefPN}) resorts to the
map $\v{n}$. The analysis will proceed through the determination of
the largest subdomains $\{\Brill'_i\}$ of invertibility of the map
$\v{n}$. We first prove that on the closure $\overline\Brill_i$ of
each domain the map is surjective on the closed unit sphere
$\overline\Ball$. Then, we determine the geometry of the ranges
$\v{n}(\Brill'_i)$, showing that they are homotopic to a solid
genus-two torus.


\begin{figure*}[t]
\includegraphics[width=.32\textwidth]{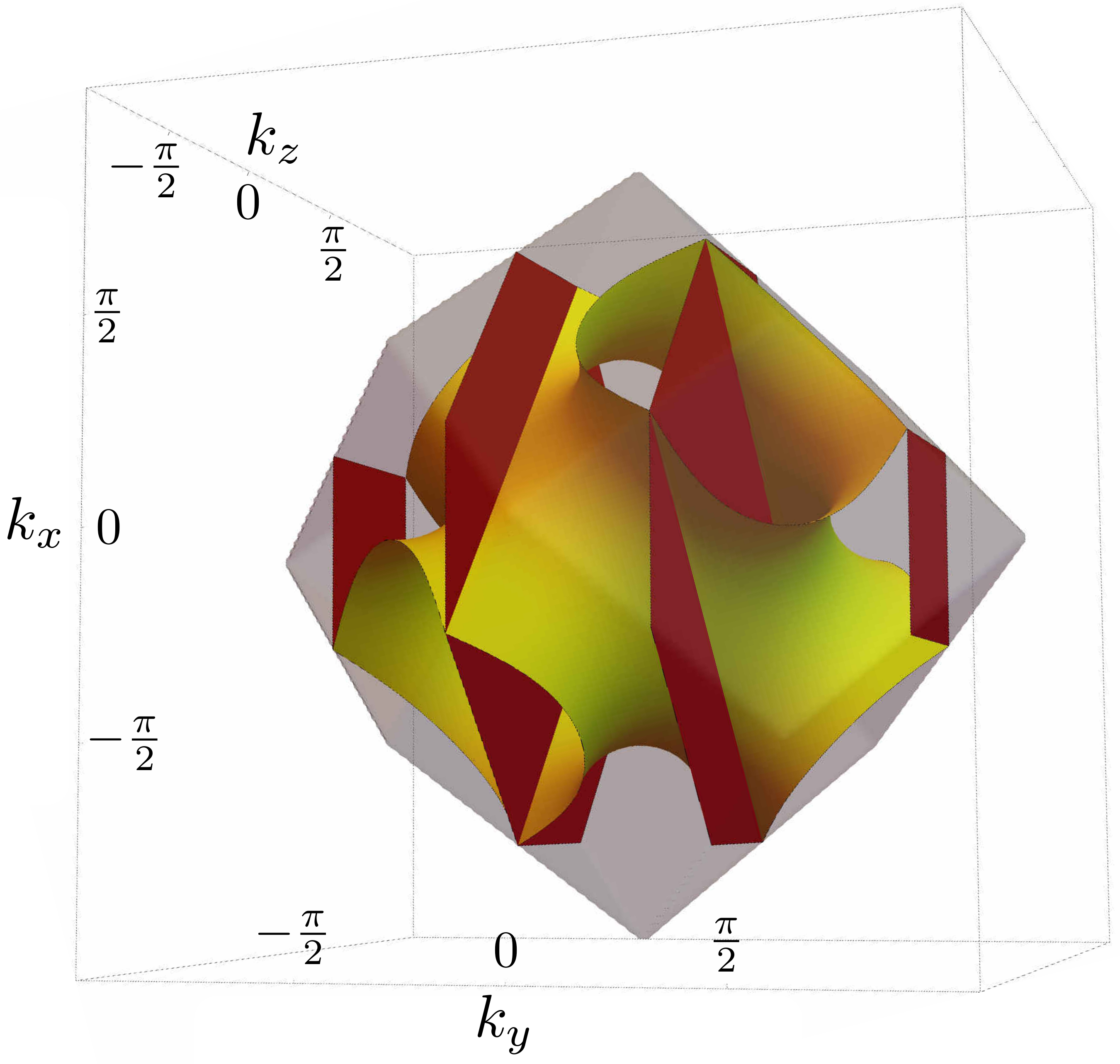}
\includegraphics[width=.32\textwidth]{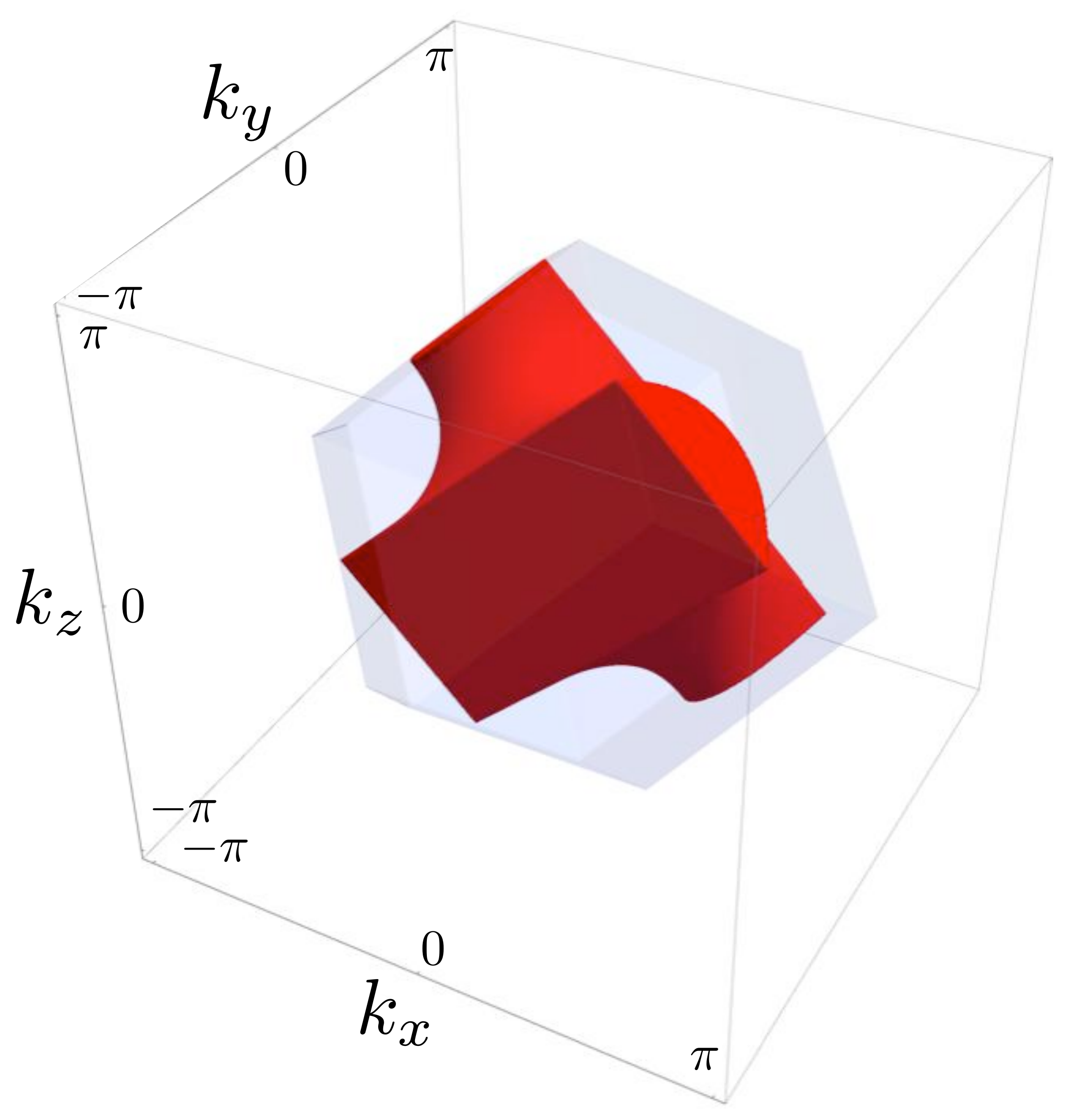}
\includegraphics[width=.32\textwidth]{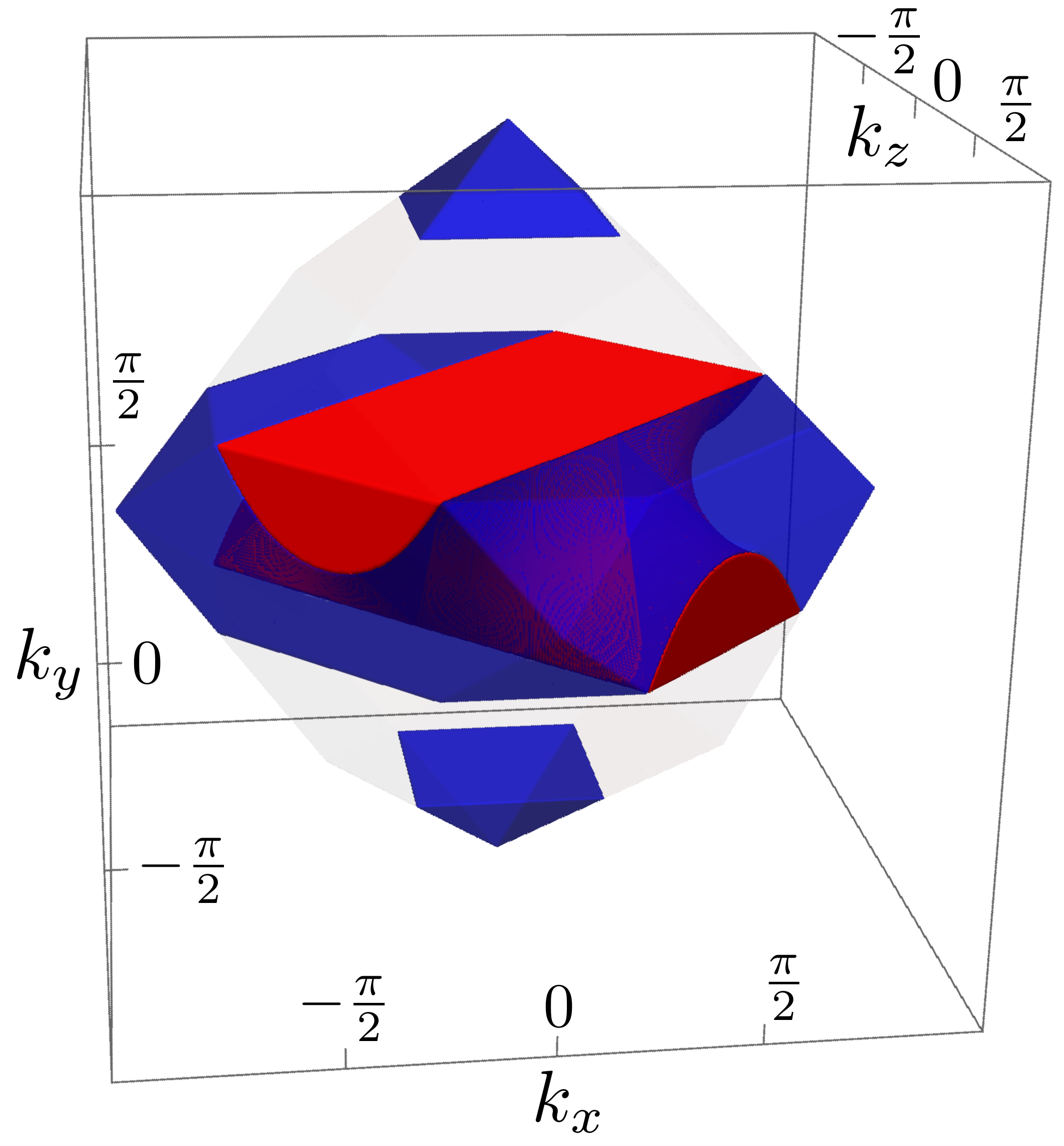}
\\
\includegraphics[width=.245\textwidth]{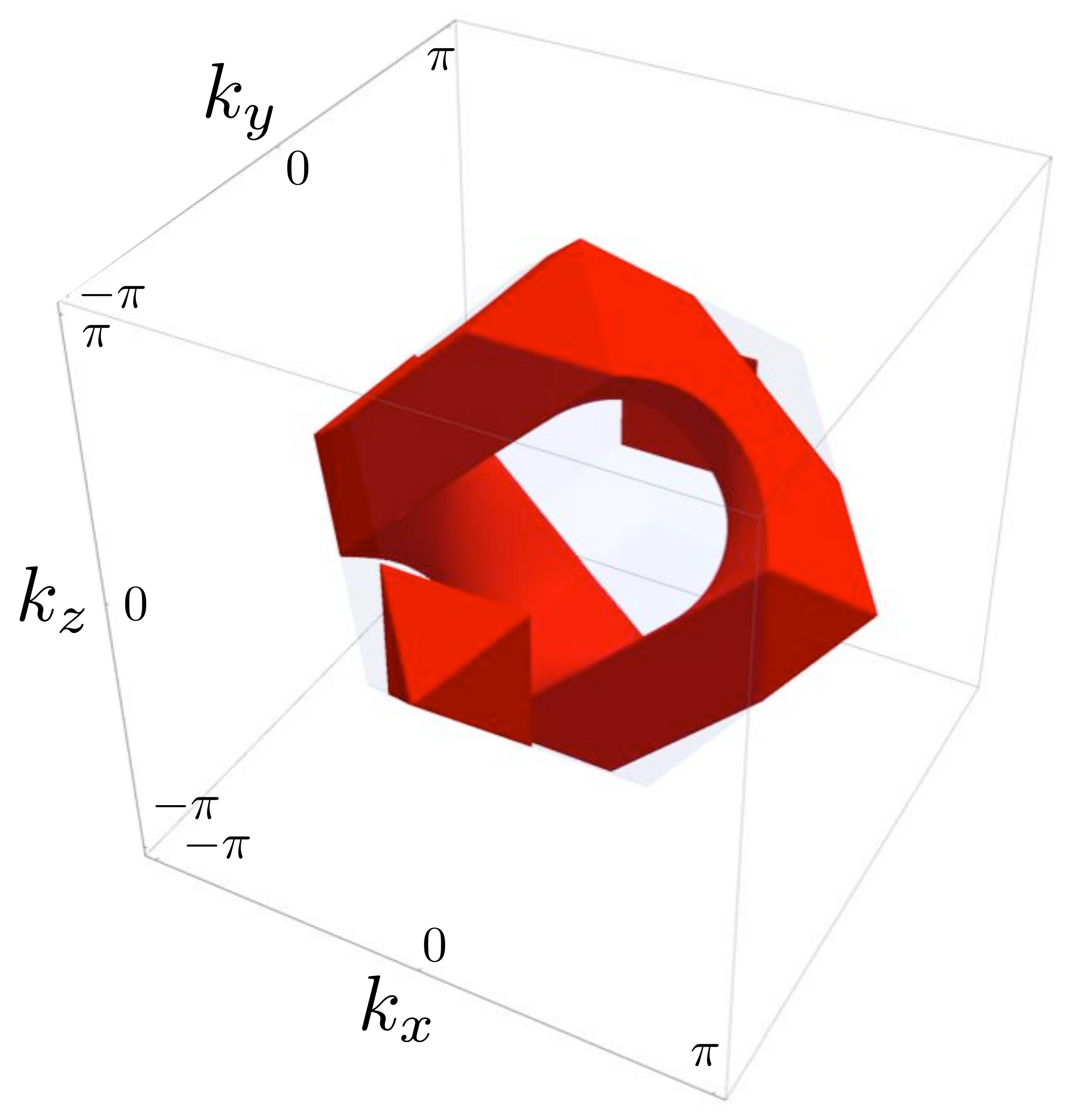}
\includegraphics[width=.245\textwidth]{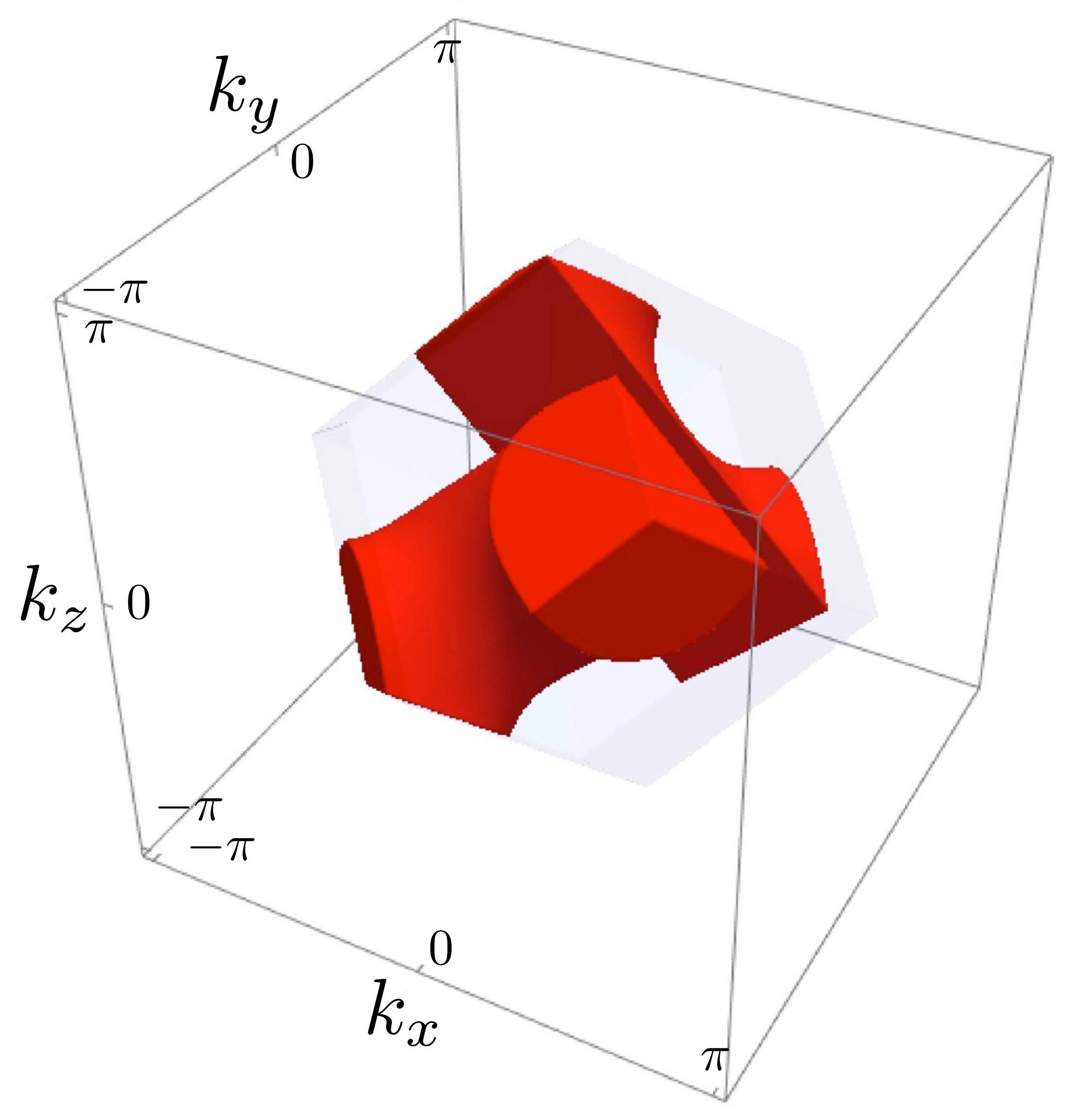}
\includegraphics[width=.245\textwidth]{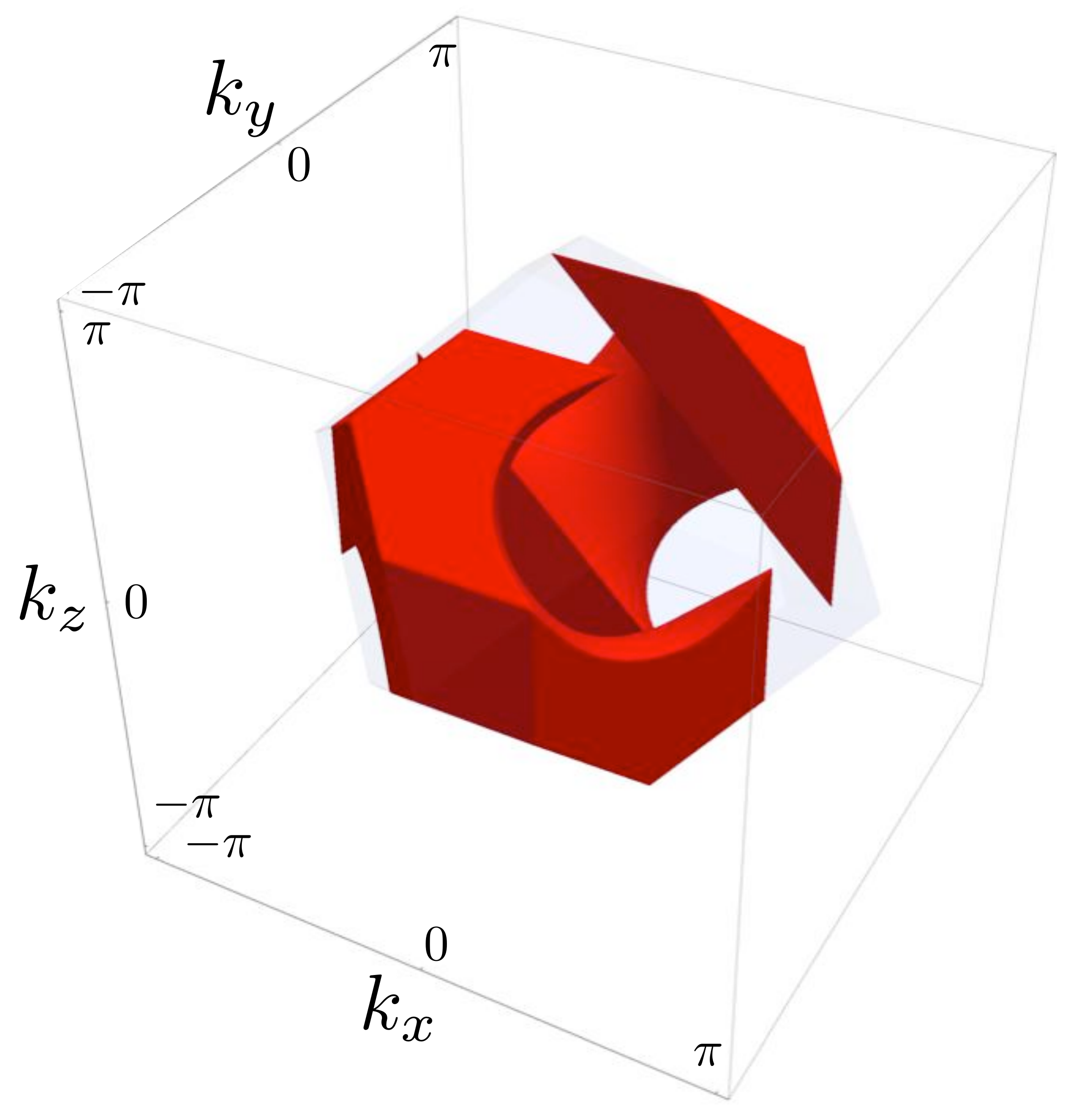}
\includegraphics[width=.245\textwidth]{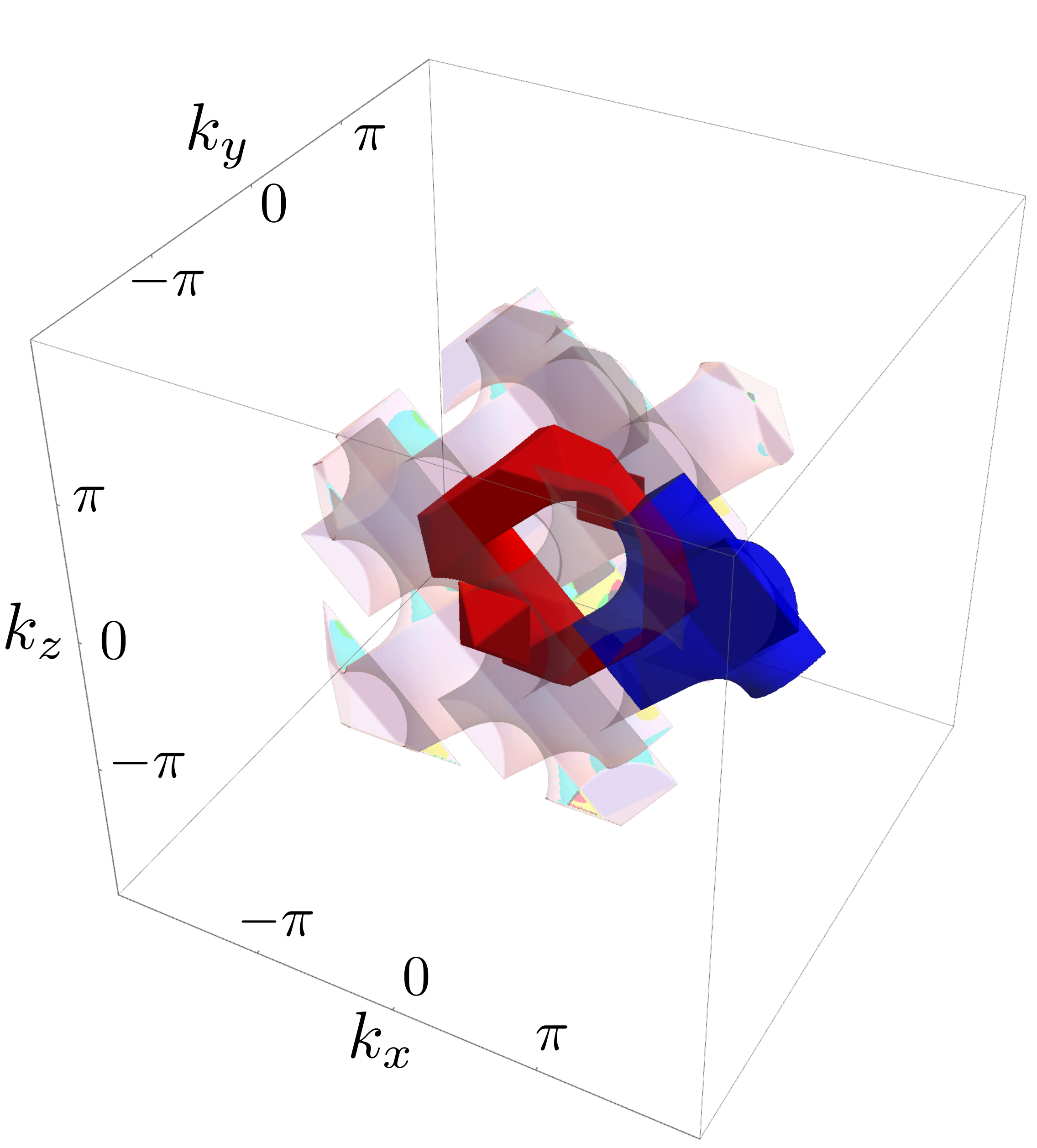}
\caption{(Colors online).
Top left figure: surfaces $\lambda(\v{k})=0$ in Eq. (\ref{eq:Slambdabis}) (yellow) and $\cos(2k_y)=0$ (red planes) inside the Brillouin zone (transparent). Top middle figure: $\Brill_0$ zone (red X-shaped). Top right figure: $\Brill_0$ (red) and $\Brill_1$ (blue). Bottom left to right: $\Brill_1$, $\Brill_2$, $\Brill_3$. Bottom right: region $\Brill_1$ represented in a properly translated Brillouin zone.\\
In this paper the Lorentz transformations are those that leave the dispersion relations of the Weyl QW invariant, and act on the Weyl spinor independently of the wave-vector. In such way they are nonlinear in $(\omega,\v{k})$ and linear over the Weyl spinor. Therefore the Lorentz group acts as a group of diffeomorphisms over the Brillouin zone $\Brill$. The four domains $\Brill_i\subset\Brill$ are Lorentz invariant  (up to a null-measure set, see Fig.  \ref{theimageofn}). More precisely a point $(\omega, \v{k})$ with $\v{k} \in \Brill_i$ and $\sin^2\omega - |\v{n}(\v{k})|^2 =0$ is mapped to a point  $(\omega', \v{k}')$ with $\sin^2\omega' - |\v{n}(\v{k}')|^2 =0$ and $\v{k}' \in \Brill_i$. Moreover, the map $\v{n}$ maps each $\Brill_i$ into the same set (up to null measure set: see Fig. \ref{theimageofn}). Since the kinematics of a wavevector $\v{k}$ depends only on the vector $\v{n}(\v{k})$, we can conclude that the $\Brill_i$ regions are kinematically equivalent and they can be interpreted as four different massless Weyl Fermions.  Because of the identification of the boundary points in  the Brillouin zone, all the $\Brill_i$ regions have the same X-shape as $\Brill_0$. This is evident in the bottom right figure, in which we see that the region $\Brill_1$ (in red), when represented in a properly translated Brillouin zone (in blue), has the same X-shape as the region $\Brill_0$. Considering the identification of the boundary points of the Brillouin zone in Fig. \ref{Brillouinzone}, one realizes that the opposite arms of the X are glued together, resulting in a solid double-torus (genus-two). This result is rigorously proved in the text where we show that the $\Brill_i$ regions are diffeomorphic to a solid ball pierced by two  arches of ellipses (Fig. \ref{theimageofn}).}
\label{fig:invariantzone}
\end{figure*}    

Let us denote by $\Brill$ the Brillouin zone of the center cubic
lattice. $\Brill$, upon a proper identification of its
boundary points (see Fig. \ref{Brillouinzone})  is a compact $3$-dimensional manifold.
The Jacobian $J_{\v{n}}(\v{k})$  of the map $\v{n}(\v{k})$ is given by
\begin{equation}\label{e:Jac}
J_{\v{n}}(\v{k}):=\det[\partial_in_j(\v{k})]=\cos(2k_y)\lambda(\v{k}),
\end{equation}
and it vanishes on the set
\begin{align}
\label{eq:Sboundary2}
\Sing &=  \Tromb \cup \Ky,\nonumber \\
\Ky &:= \{ \v{k} \in \Brill |\cos(2 k_y)=0\}, \\
 \Tromb &:=  \{ \v{k} \in \Brill | \lambda(\v{k}) = 0  \}.\nonumber
\end{align}
Since $\nabla \lambda(\v{k}) \neq 0$ for all $\v{k}$ such that $\lambda(\v{k})=0$, the implicit
function theorem guarantees that $\Tromb$ is a well defined $2$-dimensional surface. In the
following we will denote by $\{\Brill'_i\}$ ($i$ ranging in some set) the disjoint connected subsets
of $\Brill\setminus\Sing$, thus
\begin{equation}
  \Brill\setminus\Sing=\bigcup_i\Brill'_i,\quad \Brill'_i\cap\Brill'_j=\emptyset\text{ for }i\neq j.\end{equation}
For each $i$ the set $\Brill'_i$ is open and we denote as
$\overline{\Brill'_i}$ its closure and as $\partial{\Brill'_i}$ its boundary.

Now let us denote with $\overline{\Ball }\subset \mathbb{R}^3$ the closed unit-radius ball, and with
$\mathsf{S}^2$ the sphere of radius $1$ in $\mathbb{R}^3$.
Moreover, let us define the parametric curves
\begin{align}
\label{eq:Sellipsis}
\v{e}_\pm(t):=\frac{1}{\sqrt{2}}   
(      \sin(t),
    \cos(t),
    \pm\sin(t)
)^T
\end{align}
and the sets
\begin{align}
  \label{eq:Simages}
  \begin{aligned}
 \mathsf{Q}_a &:=  \Ball  \setminus (\v{e}_+(\mathsf{T}_1)\cup\v{e}_-(\mathsf{T}_2))\\
 \mathsf{Q}_b &:=  \Ball  \setminus (\v{e}_+(\mathsf{T}_2)\cup\v{e}_-(\mathsf{T}_1))\\
\mathsf{T}_1 &:= (- \tfrac{\pi}{2}, \tfrac{\pi}{2})\\
\mathsf{T}_2 &:= (-\pi,- \tfrac{\pi}{2})\cup( \tfrac{\pi}{2},\pi].   
  \end{aligned}
\end{align}
Given all the definition introduced in this section, we
  have then the following result:
\begin{lemma}
\label{lem:charactbrill}
  There are four different $\Brill'_i$ regions, determined 
 by the following conditions
\begin{align}
  \label{eq:Szones}
  \begin{aligned}
    \Brill'_0 &:=  \{\v{k}\in\Brill |\lambda(\v{k}) > 0, \cos(2 k_y) > 0\},\\
 \Brill'_1 &:=  \{\v{k}\in\Brill|\lambda(\v{k}) < 0, \cos(2 k_y) > 0\},\\
 \Brill'_2 &:=  \{\v{k}\in\Brill|\lambda(\v{k}) > 0, \cos(2 k_y) < 0\},\\
 \Brill'_3 &:=  \{\v{k}\in\Brill|\lambda(\v{k}) < 0, \cos(2 k_y) < 0 \}.
  \end{aligned}
\end{align}
For each $i$, let $\v{n}^{(i)}(\v{k})$ denote the restriction of the map
$\v{n}(\v{k})$ to the set $\Brill'_i$. Then $\v{n}^{(i)}(\v{k})$
defines a 
diffeomorphism between $\Brill'_i$ and its image
$\v{n}^{(i)}(\Brill'_i)$
and we have
\begin{align}
  \label{eq:Simages2}
  \begin{aligned}
  \v{n}^{(0)}(\Brill'_0)=  \v{n}^{(2)}(\Brill'_2)= \mathsf{Q}_a \\
\v{n}^{(1)}(\Brill'_1)=  \v{n}^{(3)}(\Brill'_3)= \mathsf{Q}_b.
  \end{aligned}
\end{align}
\end{lemma}
The proof of this result is rather involved and can be found in Appendix \ref{app:dom}.
The $\Brill'_i$ regions are plotted in Fig. \ref{fig:invariantzone} 
The most important consequence of this result is that, for each $i$,
 the set $\v{n} (\Brill'_i) $ 
(see Fig. \ref{theimageofn}) i)
 coincides with
$\Ball $ except a null measure set and ii) it is homeomorphic to a
genus two torus.


\subsection{Study of the map $\mathcal{N}$}\label{s:suitable}

\begin{figure*}[t]
\includegraphics[width=.4\textwidth]{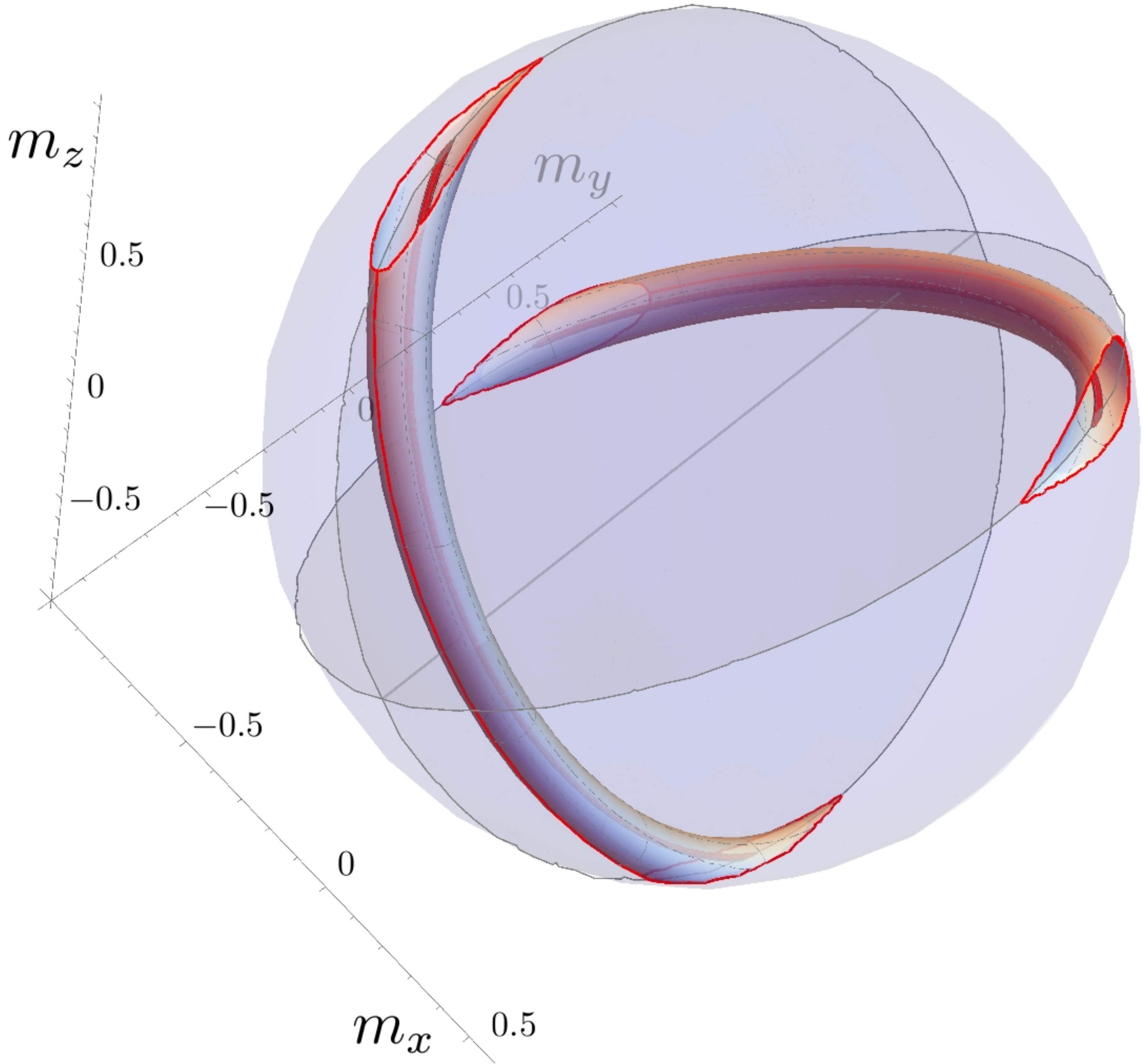}
\qquad\qquad
\includegraphics[width=.4\textwidth]{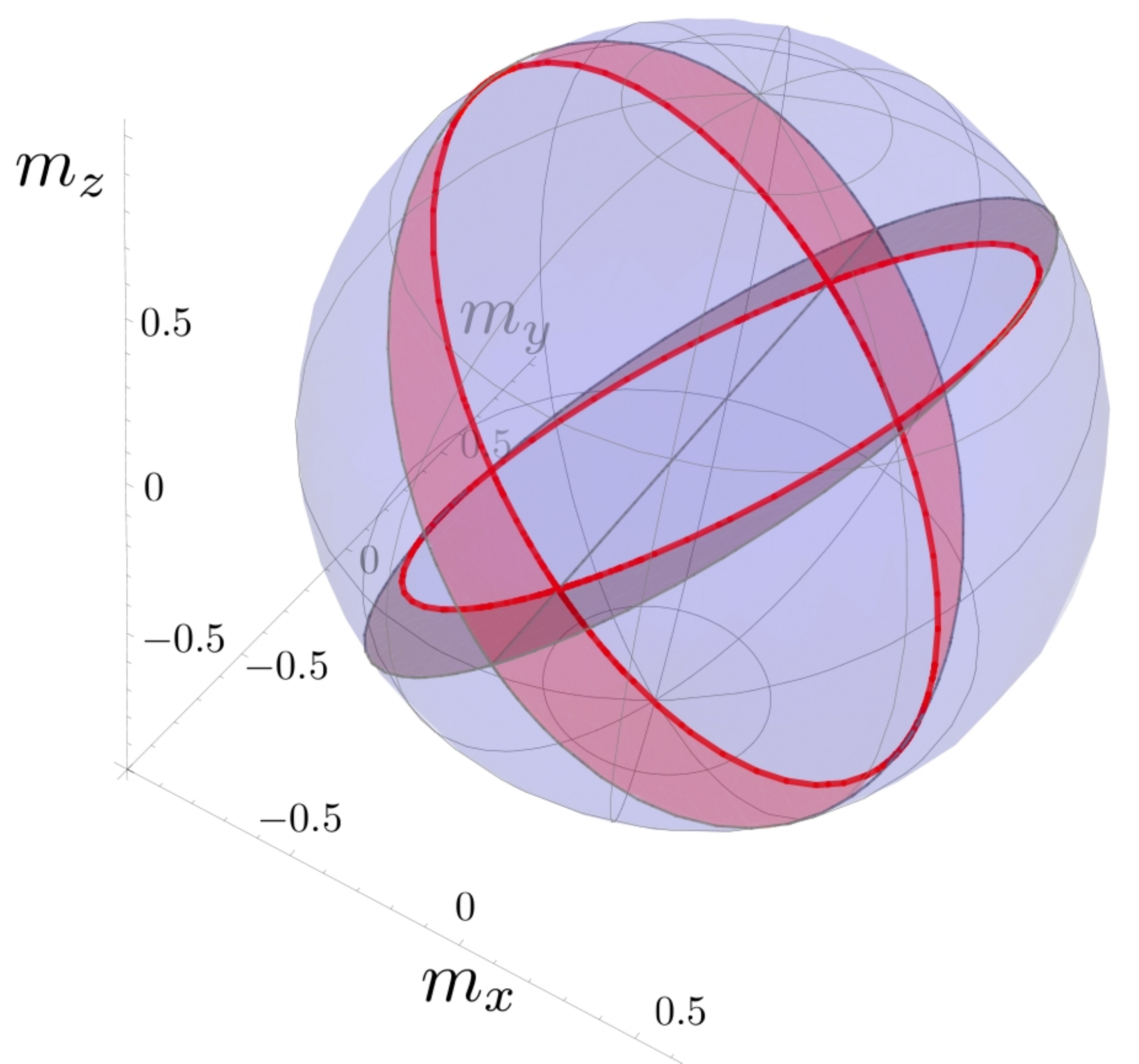}
\caption{(Colors online) Left figure: region $\mathsf{Q}_a$. 
Right figure: $\mathsf{H}$ zone in red inside the unit ball.
In the left figure, the tubes around the arches $\v{e}_+(\mathsf{T})_1$ and   $\v{e}_-(\mathsf{T})_2$ emphasize the
piercing of the ball by the one-dimensional holes along the elliptic arches $\v{e}_+(\mathsf{T})_1$ and $\v{e}_-(\mathsf{T})_2$.
The region $\mathsf{Q}_a$ is clearly homeomorphic to a solid torus of genus two. Because of this non-trivial topological feature
the set $\{(\omega , \v{m}) \text{ s.t. } |\omega| \leq \tfrac{\pi}{2}, \v{m} \in  \v{n}(\Brill_i),   \sin^2 \omega -
|\v{m}|^2 =0\}$ cannot be diffeomorphic to any Lorentz-invariant region of $\mathbb{M}^4$. However it is possible to remove from the region
$\mathsf{Q}_a$ a null-measure set such that the resulting topology is trivial. This can be done by removing the set $\mathsf{H}$ (red zones in the right figure), resulting in a star-shaped open set in $\mathbb{R}^3$.}
\label{theimageofn}
\end{figure*}

Since for all $i$ the region $\v{n}(\Brill'_i)$ has a nontrivial topology,
the set $\{(\omega , \v{m}) \text{ s.t. } |\omega| \leq
\tfrac{\pi}{2}, \v{m} \in  \v{n}(\Brill'_i),   \sin^2 \omega -
|\v{m}|^2 =0\}$ cannot be diffeomorphic to any
Lorentz-invariant region of $\mathbb{M}^4$. A possible way to change the topology of
$\v{n}(\Brill'_i)$ is to exclude the set
$\mathsf{H}\subseteq\overline\Ball$ 
(as it is shown in Fig. \ref{theimageofn}) 
of vectors $\v{m}$
satisfying the following inequalities
\begin{align}
\left\{
\begin{aligned}
&m_x=\pm m_z,\\
&2m_x^2+m_y^2\leq 1,\\
&2m_x^2+2m_y^2\geq 1.
\end{aligned}
\right.
\end{align}
Then, the set $\Ball\setminus\mathsf{H}$ is
topologically trivial and we have $\Ball\setminus\mathsf{H} \subset
\v{n}(\Brill'_i)$ for all $i$. Let us now consider the function
$\mathcal{N}: (\omega, \v{m}) \mapsto (p_0, \v{p})=g(\omega , \v{m}) (\sin \omega,
\v{m}) $ restricted to the set 
\begin{align}
  \label{eq:SthesetN}
  \mathsf{N}:=\{ (\omega, \v{m}) \text{ s.t }
\v{m} \in \Ball\setminus\mathsf{H},   |\omega| \leq
\tfrac{\pi}{2},  \sin^2 \omega -
|\v{m}|^2 =0 \}.
\end{align}
As shown in Appendix \ref{app:func}, it is possible to define the function
$g(\omega , \v{m})$ is such that $\mathcal{N}$
defines a diffeomorphism between $\mathsf{N}$
and the null mass-shell
\begin{align}
 \mathsf{\Gamma}_0 := \{ p\in \mathbb{M}^4,  \text{ s.t. }  p^\mu p_\mu = 0 \}. 
\end{align}
and that its Jacobian matrix at tho origin is $0$, i.e
\begin{align}
\label{eq:Snear0}
J_\mathcal{N}(\v{0}) = I.
\end{align}

Finally, for each $ i $ we denote by $\Brill_i$ 
the counter-image of the set $\mathsf{U} \setminus  \mathsf{H}$
under the map $\v{n}^{(i)}$ and by $\mathcal{D}^{(i)}$
the composition
\begin{align}
\label{eq:Stheultimatedeformation}
 & \mathcal{D}^{(i)} :  \mathsf{\Sigma}_i \to \mathsf{\Gamma}_0 \qquad
  \mathcal{D}^{(i)} := \mathcal{N} \circ \mathcal{P}^{(i)} \\
&  \begin{aligned}
&\mathcal{P}^{(i)} : \mathsf{\Sigma}_i  \to  \mathsf{N}  \qquad \mathcal{P}^{(i)} : 
\begin{pmatrix}
  \omega\\
\v{k}
\end{pmatrix}
 \mapsto 
\begin{pmatrix}
  \omega\\
\v{n}^{(i)}(\v{k})
\end{pmatrix}\\
&\mathcal{N} :  \mathsf{N} \to \mathsf{\Gamma}_0
\qquad
\mathcal{N} : 
\begin{pmatrix}
  \omega\\
\v{m}
\end{pmatrix}
 \mapsto 
g(\omega, \v{m})
\begin{pmatrix}
  \sin \omega\\
\v{m}
\end{pmatrix}  \\
&\mathsf{\Sigma}_i := \{ (\omega , \v{k}) \text{ s.t. }   \v{k} \in
\Brill_i , \sin^2\omega - |\v{k}|^2 = 0 \}.
  \end{aligned} 
\nonumber
  \end{align}
For each $ i $, the map $\mathcal{D}^{(i)}$ is an analitic diffeomorphism
between the region $ \mathsf{\Sigma}_i$ and the Lorentz invariant set
$\mathsf{\Gamma}_0$ which satisfies the condition
$J_{\mathcal{D}_i}(\bvec{0})= I$. Then the composition
\begin{align}
  \label{eq:Stheultimatelorentz}
  \mathcal{L}^{(i)}_{\beta}: \mathsf{\Sigma}_i \to \mathsf{\Sigma}_i \qquad 
 \mathcal{L}^{(i)}_{\beta}  := \mathcal{D}^{-1} \circ L_\beta  \circ  \mathcal{D}
\end{align}
is  a well defined nonlinear representation of the Lorentz group on
the set  $\mathsf{\Sigma}_i$.
Since the union of the $\Brill_i$ sets
coincides with the whole (up to a null measure set) Brillouin zone, we
have that the collection of the maps
  $ \mathcal{L}^{(i)}_{\beta}$ provide a notion of Lorentz
  transformation for any (up to a null measure set) solution of the
  Weyl QW dynamics.

\section{Proof of Lemma \ref{lem:charactbrill}}\label{app:dom}
In this section will give the proofs of the results contained in Lemma
\ref{lem:charactbrill}. Since the proof is quite involved, we split it
into several pieces. Let us begin by  defining the sets
\begin{align}
  \mathsf{Q}' :=  \overline{\Ball } \setminus \mathsf{R} \qquad
\mathsf{R} :=
 \mathsf{S}^2 \cup
 \mathsf{E}_+ \cup \mathsf{E}_- \,.
\end{align}
Obviously $\mathsf{Q}'$ is open and connected, with
 $\overline{\mathsf{Q}'}=\overline{\Ball }$ and
$\partial \mathsf{Q}' = \mathsf{R}$.
   We now prove some useful  properties of the map $\v n $.
\begin{sublemma}
\label{lem:diffeo}
  Let
$\v n _i$ denote the restriction of the map $\v n $ to $\Brill'_i$.
 Then
  for each $i$ we have that
 $\v n _i$ is a diffeomorphism between
    $\Brill'_i$ and $\v n ({\Brill'_i})$
\end{sublemma}
\Proof Since by definition $\v{k} \in \Brill'_i \Rightarrow \v{k} \not \in \Sing $ we have $J_{\v n
}(\v{k}) \neq 0$ for all $\v{k} \in \Brill'_i$. Since $\Brill'_i$ is connected and $\v n $ is
analytical, we have the thesis.  \qed
\begin{sublemma}
\label{sch:inclusions}
 We have the following inclusions:
 \begin{enumerate}
 \item \label{item:incleasy} $\v n (\overline{\Brill'_i}) \subseteq \overline{\Ball }$,
 \item \label{item:inclhard}$\partial \v n ({\Brill'_i}) \subseteq \mathsf{R}$.
\end{enumerate}
\end{sublemma}
\Proof
 Let us start with the proof of item \ref{item:incleasy}.  By explicit computation we have $ |\v n (\v{k})|^2 =
1- \lambda^2(\v{k})\leq 1$ which implies that the image of
$\v n $ is contained in $\overline\Ball $. 

We now prove item \ref{item:inclhard}.  Thanks to Lemma
\ref{lem:diffeo} we have that $\v n ({\Brill'_i})$ is
open. On the other hand, since $\v n $ is continuous and
$\overline{\Brill'_i}$ is compact, we have that
$\v n (\overline{\Brill'_i})$ is compact and then it is closed.
Then the trivial inclusion 
 $\v n ({\Brill'_i}) \subseteq \v n (\overline{\Brill'_i})$
 implies
 $\overline{\v n ({\Brill'_i})} \subseteq
 \v n (\overline{\Brill'_i})$.
By definition we have 
$\overline{\v n ({\Brill'_i})} =
\v n ({\Brill'_i}) \cup \partial \v n ({\Brill'_i})  $
with $
\v n ({\Brill'_i}) \cap \partial \v n ({\Brill'_i}) =
\emptyset  $ and 
$ \v n (\overline{\Brill'_i} ) =
\v n ({\Brill'_i}) \cup \v n (\partial \Brill'_i)$.
Then the inclusion
 $\overline{\v n ({\Brill'_i})} \subseteq
 \v n (\overline{\Brill'_i})$ implies
$ \partial \v n ({\Brill'_i}) \subseteq \v n (\partial
\Brill'_i)$.
Since $\partial \Brill'_i \subseteq \Sing$ we have
$\v n (\partial \Brill'_i) \subseteq \v n (\Sing) $.
One can then verify by direct computation that $ \v n (\Sing)
\subseteq \mathsf{R}$ thus proving the thesis.
\qed

We now recall a result of basic topology which will be useful in the following.
\begin{sublemma}
\label{sch:topo}
Let ${A}$ and $ {B}$ be open sets such that $\overline A\subset \overline B$.  Then there exists a
point $p$ such that $p\in\mathrm{int}\overline B$ and $p \not \in\overline A $.
\end{sublemma}
\Proof Let us suppose that $ B\subseteq\overline A$.  Since
$ B$ is open and $\overline A $ is closed, we have $ \overline{B}
\subseteq\overline A$ which contradicts the hypothesis.  \qed

The following result will be of crucial importance.
\begin{sublemma}
\label{lem:cruciallemma}
  The following identity holds: 
  \begin{align}
    \label{eq:Scrucialidentity}
\overline{\v n (\Brill'_i)} = \overline{\Ball }.    
  \end{align}
\end{sublemma}
\Proof
First we prove the easiest inclusion
$\overline{\v n (\Brill'_i)} \subseteq \overline{\Ball }$.
From item \ref{item:incleasy} of Sublemma \ref{sch:inclusions} 
we have that
$\v n (\Brill'_i) \subseteq \v n (\overline{\Brill'_i})
\subseteq \overline{\Ball }$ (the first inclusion is trivial).
Reminding that $\v n (\Brill'_i)$ is open we have
$\overline{\v n (\Brill'_i)} \subseteq \overline{\Ball }$.

We now prove that
$ \overline{\Ball } \subseteq \overline{\v n (\Brill'_i)} $.  By
contradiction, let us suppose that the strict inclusion
$\overline{\v n (\Brill'_i)} \subset \overline{\Ball } $ holds.  Then,
thanks to Sublemma \ref{sch:topo}, we find $p \in \Ball $ such that
$p \not \in \overline{\v n (\Brill'_i)} $.  Moreover we can find an
open neighborhood $\mathsf{N}$ of $p$ such that
$\mathsf{N} \cap \overline{\v n (\Brill'_i)} = \emptyset$ and then
without loss of generality we can suppose that $p \in
\mathsf{Q}'$.
Since $\mathsf{R}$ has no interior points, $\v{n}(\Brill'_i)$ cannot be
included in $\mathsf{R}$, whence $\v n (\Brill'_i) \cap \mathsf{Q}'$ is
not empty. Let us now fix a point
$q \in \v n (\Brill'_i) \cap \mathsf{Q}'$. Then, for any continuous
path $\gamma$ connecting $p$ and $q$ there exist $t'$ such that
$\gamma(t') \in \partial \v n (\Brill'_i) $.  From item
\ref{item:inclhard} of Sublemma \ref{sch:inclusions} we have
$\gamma(t') \in \mathsf{R}$.  Since this conclusion contradicts the
fact that $\mathsf{Q}'$ is connected, we have proved the thesis.  \qed

As a consequence we have

\begin{corollary}
\label{cor:whatweneed}
The following inclusion holds 
$\mathsf{Q}' \subseteq \v n (\Brill'_i)$.
\end{corollary}
\Proof
From Lemma \ref{lem:cruciallemma} we have
$\mathsf{Q}' \cup \mathsf{R} = \v n (\Brill'_i) \cup \partial
\v n (\Brill'_i)$.
Reminding that 
$\mathsf{Q}' \cap \mathsf{R} = \emptyset = 
\v n (\Brill'_i) \cap \partial
\v n (\Brill'_i)$ and the inclusion $\partial
\v n (\Brill'_i) \subseteq \mathsf{R}$, proved in Sublemma \ref{sch:inclusions},
we have the thesis.
\qed

We now turn our attention to the regions
$\Brill'_i$. Our first objective is to determine how many different
$\Brill'_i$ regions are. The answer is provided by the following result.
\begin{sublemma}
  The regions $\Brill'_i$  are in one-to-one correspondence
with the solution of the equation $|\lambda(\v{k})|^2 =1 $.
\end{sublemma}
\Proof 
We proved that the map
$\v{n}_i$ defines a diffeomorphism between $\Brill'_i$ 
and the set $\v{n}(\Brill'_i) \subseteq \overline{P}$
which includes the origin. Then, for each $\Brill'_i$ there exist 
a point $\v{k}$ such that $\v{n}(\v{k}) = 0$ and it is unique.
Since $\v{n}(\v{k}) = 0$ if and only if
$|\v{n}(\v{k})|^2 = |\lambda(\v{k})|^2 -1= 0$ we have the
thesis. \qed
Thanks to this result it is sufficient to find the solutions of
$|\lambda(\v{k})|^2 =1 $ in the Brillouin zone. One can easily check
that there are only $4$ solutions and then $4$ different regions
$\Brill'_0 , \dots, \Brill'_3$.

We can now prove Eq. \eqref{eq:Szones} of Lemma \ref{lem:charactbrill}.
\begin{sublemma}
The region $\Brill'_i$ are given by
\begin{align}
  \label{eq:Scharacteregion}
  \begin{aligned}
   \Brill'_0 &:=  \{\v{k}\in\Brill |\lambda(\v{k}) > 0, \cos(2 k_y) > 0\},\\
 \Brill'_1 &:=  \{\v{k}\in\Brill|\lambda(\v{k}) < 0, \cos(2 k_y) > 0\},\\
 \Brill'_2 &:=  \{\v{k}\in\Brill|\lambda(\v{k}) > 0, \cos(2 k_y) < 0\},\\
 \Brill'_3 &:=  \{\v{k}\in\Brill|\lambda(\v{k}) < 0, \cos(2 k_y) < 0 \}.
  \end{aligned}
\end{align}  
\end{sublemma}
\Proof
Let us denote with $\tilde {\Brill}_i$
the regions defined by the right hand sides of Eq.
\ref{eq:Scharacteregion}.
One can immediately see that:
i) the $\tilde {\Brill}_i$ are open sets,
ii) the $\tilde {\Brill}_i$ are mutually disjoint
and iii) the union of the $\tilde {\Brill}_i$ is the union of the 
$\Brill'_i$.
We now prove that for all $\tilde {\Brill}_i$
there exist a unique $\Brill'_j$ such that
$\tilde {\Brill}_i \subseteq \Brill'_j$.
This fact, together with the previous properties of the 
$\tilde {\Brill}_i$ gives
$\tilde {\Brill}_i = \Brill'_i$

Clearly for all $\tilde {\Brill}_i$ we must have
$\tilde {\Brill}_i \subseteq \Brill'_{j_1} \cup \dots \cup
\Brill'_{j_k}$ for some $k \geq 1$
Let us suppose then that there exist 
$\tilde {\Brill}_i$
such that
$\tilde {\Brill}_i \subseteq \Brill'_{j_1} \cup \dots \cup
\Brill'_{j_k}$
with $k$ strictly greater than $1$. 
Since we have as many
$\tilde {\Brill}_i$ as $\Brill'_i$, there must exist a
$\Brill'_i$ and two points $\v{k}_a \in \tilde {\Brill}_a $
and  $\v{k}_b \in \tilde {\Brill}_a $ such that
$ \v{k}_a ,\v{k}_b \in \Brill'_i$. Since
$\Brill'_i $ is connected 
there must exist a path connecting 
$ \v{k}_a$ and  $\v{k}_b$ that entirely lies within 
$\Brill'_i$. On the other hand, since  the $\tilde {\Brill}_i$
are disjoint,  this path would cross the border of the  $\tilde
{\Brill}_a$ but this contradicts the fact that the border of the  
$\tilde{\Brill}_a$ are not included in $\Brill'_i$.
\qed

Finally, we can give the complete characterization of the sets
$ \v{n}(\Brill'_i)$.
From  Corollary \ref{cor:whatweneed} we have the inclusion
 $\mathsf{Q}' \subseteq \v{n}(\Brill'_i)$.
Since $|\v{n}(\v{k})|=1 \iff \lambda(\v{k}) =0$,
we know that the $S^2 \not \subset \v{n}(\Brill'_i)$.
It is easy to check that also the points 
$ p_{\pm} :=  ( 0, \pm\tfrac{\sqrt2}{2} , 0  )$
are not included in the set
 $\v{n}(\Brill'_i)$.
For any region $\v{n}(\Brill'_i)$, 
we will determine which ones of the $8$ open arches defined as
\begin{align}
  \begin{aligned}
&\mathsf{E}^{\pm}_j :=\v{e}_{\pm} (L_j) \\
&L_1 := (0, \tfrac{\pi}{2})\qquad
L_2 := (\tfrac{\pi}{2} ,  \pi )\\
&L_3 := (-\tfrac{\pi}{2} ,  0)\qquad
L_4 :=  ( -\pi ,  -\frac{\pi}{2})
  \end{aligned}
\end{align}
are included in $\v{n}(\Brill'_i)$.

Let us consider the sets  $\mathsf{E}^{+}_j $.
If we for some $t$ we have 
$  \v{n}(\v{k}) = \v{e}_+(t)$ and $ \cos(2k_y) \neq 0$
then it must be
\begin{align}
  \label{eq:Sconditionkx}
  k_x = \tfrac{\pi}{4} + n \tfrac{\pi}{2}
\qquad  k_z = k_x + m \pi
\end{align}
for $n$ and $m$ integers.
Eq. \eqref{eq:Sconditionkx} then implies 
\begin{align}
\label{eq:Sconditionlambda}
\lambda (\v{k}) = (-1)^{m} \frac12 (\cos(k_y) - \sin(k_y)).
\end{align}
From Eq. \eqref{eq:Sconditionlambda}
we have
\begin{align}
  \label{eq:S12}
  \lambda(\v{k}) > 0 \Rightarrow 
\left \{
  \begin{aligned}
&m \mbox{ even } \wedge \minus
  \tfrac{3}{4} \pi < k_y <\tfrac{1}{4} \pi
\\
&m \mbox{ odd } \wedge 
  \tfrac{1}{4} \pi < k_y <\tfrac{5}{4} \pi
  \end{aligned}
\right.
\end{align}
Then, if we assume  $\v{k} \in \Brill'_0$ we must have
\begin{align}
\label{allowedsetsB0}
\begin{aligned}
  m \mbox{ even } , \minus
  \tfrac{1}{4} \pi < k_y <\tfrac{1}{4} \pi  \\
  m \mbox{ odd } , 
  \tfrac{3}{4} \pi < k_y <\tfrac{5}{4} \pi  
\end{aligned}
\end{align}
However,  since the two sets of $\v{k}$ are related by a translation of
$(0,l \pi , l \pi ), l \in \mathbb{Z}$ they actually represent the
same set in the Brillouin zone. 
So it suffices to consider just the
first set in Eq. \eqref{allowedsetsB0}, that can be written as
\begin{align}
\label{eq:Sallowesets2B0}
\begin{aligned} 
  &k_x =  \tfrac{1}{4} \pi + n \tfrac{1}{2} \pi \qquad \tfrac{1}{4} \pi < k_y <\tfrac{1}{4} \pi\\
  &k_z = \tfrac{1}{4} \pi + n \tfrac{1}{2} \pi + m \pi = \tfrac{1}{4}
  \pi + n \tfrac{1}{2} \pi
\end{aligned}
\end{align}
where we used Eq. \eqref{eq:Sconditionkx}
and in the second equality in the second line of
Eq. \eqref{eq:Sallowesets2B0}
we used the hypothesis that $m$ is even.
Using again the fact that we identify points 
 related by a translation of
$(l \pi, 0 , l \pi ), l \in \mathbb{Z}$ we find just two inequivalent
sets
\begin{align}
\mathsf{Z}_1 := 
\left \{
\begin{aligned} 
  &k_x =  \tfrac{1}{4} \pi \\
  &\tfrac{1}{4} \pi < k_y <\tfrac{1}{4} \pi \\
&k_z = \tfrac{1}{4} \pi 
\end{aligned}
\right.
\quad
\mathsf{Z}_2 := 
\left \{
\begin{aligned} 
  &k_x =  -\tfrac{1}{4} \pi \\
  &\tfrac{1}{4} \pi < k_y <\tfrac{1}{4} \pi \\
&k_z = -\tfrac{1}{4} \pi 
\end{aligned}
\right.
.
\nonumber
\end{align}
It is now easy to show that 
the images of this two sets under the map $\v{n}$
are $\v{n}(\mathsf{Z}_1) = \mathsf{E}^{+}_{2}$ and 
$\v{n}(\mathsf{Z}_2) = \mathsf{E}^{+}_{4}$.
By applying an analogous line of reasoning 
one can prove  all the
following  inclusions
\begin{align}
  \label{eq:Sallthearches}
  \begin{aligned}
   (\mathsf{E}^{+}_{2}\cup \mathsf{E}^{+}_{4} \cup \mathsf{E}^{-}_{1}\cup \mathsf{E}^{-}_{3})
 \subset
   \v{n}(\Brill'_0) \\
 (\mathsf{E}^{+}_{1}\cup \mathsf{E}^{+}_{3} \cup \mathsf{E}^{-}_{2}\cup \mathsf{E}^{-}_{4} )
\not\subset
\v{n}(\Brill'_0) \\
   (  \mathsf{E}^{+}_{2}\cup \mathsf{E}^{+}_{4} \cup \mathsf{E}^{-}_{1}\cup \mathsf{E}^{-}_{3} )
\subset
   \v{n}(\Brill'_2) \\
 (\mathsf{E}^{+}_{1}\cup \mathsf{E}^{+}_{3} \cup \mathsf{E}^{-}_{2}\cup \mathsf{E}^{-}_{4} )
\not\subset
\v{n}(\Brill'_2) \\
  ( \mathsf{E}^{+}_{1}\cup \mathsf{E}^{+}_{3} \cup \mathsf{E}^{-}_{2}\cup \mathsf{E}^{-}_{4} )
\subset
   \v{n}(\Brill'_1) \\
    ( \mathsf{E}^{+}_{2}\cup \mathsf{E}^{+}_{4} \cup \mathsf{E}^{-}_{1}\cup \mathsf{E}^{-}_{3} )
\not\subset
   \v{n}(\Brill'_1) \\
   (  \mathsf{E}^{+}_{1}\cup \mathsf{E}^{+}_{3} \cup \mathsf{E}^{-}_{2}\cup \mathsf{E}^{-}_{4} )
\subset
   \v{n}(\Brill'_3) \\
(\mathsf{E}^{+}_{2}\cup \mathsf{E}^{+}_{4} \cup \mathsf{E}^{-}_{1}\cup \mathsf{E}^{-}_{3} )
\not\subset
   \v{n}(\Brill'_3) .
  \end{aligned}
\end{align}
This result completes the proof of Eq. \eqref{eq:Simages2}
of Lemma \ref{lem:charactbrill}.

\section{The function $g(\omega,\v{m})$}\label{app:func}
In this section we now show how it is possible to define a function $g(\omega , \v{m})$ such that the map $\mathcal{N}$ defines a diffeomorphism between $\mathsf{N}$ and the null mass-shell $ \mathsf{\Gamma}_0 $. The purpose of the following construction is to reduce the region $\mathsf{N}$ to a star-shaped region $\tilde{\mathsf N}$ by removing a null measure region, and to define the function $g(\omega , \v{m})$ in such a way that the map $\mathcal N$ is invertible on $\tilde{\mathsf N}$. Since multiplication by $g(\omega,\v m)$ rescales the four vector $(\sin\omega,\v m)$ without affecting its direction, in order to have an invertible map $\mathcal N$ it is sufficient to ensure that $g(\omega,\v m)$ is {\em radially monotonic} versus $\v m$, namely $g(\omega,r\v m_0)$ must be monotonic versus $r$. 

Let us denote $\mathsf{E}_\pm$ the ellipses defined by the parametric
equations \eqref{eq:Sellipsis}. We define the polynomials
 \begin{align}
   \label{eq:Spoly}
   \begin{aligned}
 h_\mathsf{U}(r,\theta, \phi) :=& 1- r^2 , \\
 h_\mathsf{E}(r,\theta, \phi) :=& (\cos^2(\phi) - \sin^2(\phi))^2 + \\
&+ (\tfrac{1}{2} - r^2 (1- \cos^2(\theta) \sin^2(\phi)))^2 ,   
   \end{aligned}
  \end{align}
where we used the spherical coordinates
$m_x = r \cos\theta \cos\phi$,
$m_y = r \sin\theta $,
$m_z = r \cos\theta \sin\phi$.
Clearly we have $h _\mathsf{U} (\v{m}),h_\mathsf{E} (\v{m}) > 0$ $\forall \v{n} \in $, 
$h_\mathsf{U} (\v{m}) = 0 \Leftrightarrow \v{m} \in  \mathsf{S}^2 $, $h_\mathsf{E} (\v{m}) = 0 \Leftrightarrow \v{m} \in  \mathsf{E}_+ \cup
 \mathsf{E}_-$ and $h_\mathsf{U},h_\mathsf{E}$ are analytic on $\mathsf{U} \setminus \mathsf{H}$. Since $\mathsf{U} \setminus \mathsf{H}$ is star-shaped
we can define
\begin{align}
  \label{eq:Stherelevantg}
  \tilde{g}(r,\theta,\phi) := r \int_0 ^r \!\!\!\! ds \,\,\left(
  \frac{1}{h_\mathsf{U}(s, \theta, \phi)} + \frac{1}{h_\mathsf{E}(s, \theta, \phi)} \right).
\end{align}
The condition $h _\mathsf{U} (\v{m}),h_\mathsf{E} (\v{m}) > 0$
$\forall \v{n} \in \mathsf{U} \setminus \mathsf{H}$ implies that
 the function
$\tilde{g}(r,\theta,\phi)$ is radially monotonic on
$\mathsf{U} \setminus \mathsf{H}$.
Since $h _\mathsf{U} (\v{m})$ are $h_\mathsf{E} (\v{m})$
are analytic on $\mathsf{U} \setminus \mathsf{H}$ we have that
$\tilde{g}(r,\theta,\phi)$ is analytic on
$ (\mathsf{U} \setminus \mathsf{H})\setminus \v{0}$.  Moreover,
since $\tilde{g}(r,\theta,\phi)$ is even in $r$
we have that  $\tilde{g}(\v{m})$  is analytic on the whole domain
$\mathsf{U} \setminus \mathsf{H}$.
Finally it is easy to check that 
$g(\v{m})$ diverges to $+ \infty$ as 
$\v{m}$ approaches the boundary of $\mathsf{U} \setminus \mathsf{H}$, and that  $(\nabla g)(0) = \v{0}$.
Let us define  
\begin{align}
  \label{eq:Sfinalgmap}
  g(\v{m}) := \tilde{g}(\v{m}) + 1. 
\end{align}
We now  check that, with this definition of the map $g(\v{m})$,
the map  $\mathcal{N}$
defines an analytic diffeomorphism between 
 $\mathsf{U} \setminus \mathsf{H}$ and $ \mathsf{\Gamma}_0 $
with the property $J_{\mathcal{N}}(\v{0}) = I$.
Clearly $\mathcal{N}$ is analytic in $\mathsf{N}$ so we just need to prove
that it gives a bijection between
 $\mathsf{N}$ and $ \mathsf{\Gamma}_0 $.
Let us fix a versor $\v{j}$ in $\mathbb{R}^3$. 
Then, in the $\v{j}$ direction we have
\begin{align}
  \label{eq:SultimateNmap}
\mathcal{N} (\omega,r\v j) =
  g(r, \theta_{\v{j}}, \phi_{\v{j}} )
\begin{pmatrix}
\sin (\omega) \\
r\v j
\end{pmatrix}.
\end{align}
Since $g(r, \theta_{\v{j}}, \phi_{\v{j}} ) $ is monotone we clearly
have that $\mathcal{N}$ is injective.
We now prove the  surjectivity of $\mathcal N'(\omega,r):= 
\mathcal{N}(\omega,r\v j)$ on the set $\mathsf{K} := \{ ( p_0 ,p_1 ) \in \mathbb{R}^2 \text{
   s.t. }  p_{0}^{2}-p_{1}^{2} = 0  \} $.
Let us fix a point $p=(p_0,p_1) \in \mathsf{K}$.
Since $g(r, \theta_{\v{j}}, \phi_{\v{j}} ) $ is monotone and
surjective on $[1, +\infty)$ we can find a value $r_p$ such that
$r_p g(r_p, \theta_{\v{j}}, \phi_{\v{j}} ) = p_1$.
Clearly, since $|r_p| < 1$, also the equation 
$\sin^2(\omega_p) = |r_p|^2$ can be solved and then
$\mathcal{N}'(\omega_p,r_p) = (p_0,p_1) $.
Since the surjectivity of 
$\mathcal{N}$ holds for any direction $\v{j}$,
we have that 
$\mathcal{N}$ is a diffeomorphism between 
$\mathsf{N}$ and $\mathsf{\Gamma}_0$.
Finally, 
since
$g (\v{0})=1 $ and 
 $\nabla g (\v{0}) = \v{0}$, we have
that the Jacobian of the map 
$\mathcal{N}$ is the identity, which proves Eq. \eqref{eq:Snear0}.
Obviously, our choice of the map $g(\v{m})$ is not
unique.


\begin{thebibliography}{27}%
\makeatletter
\providecommand \@ifxundefined [1]{%
 \@ifx{#1\undefined}
}%
\providecommand \@ifnum [1]{%
 \ifnum #1\expandafter \@firstoftwo
 \else \expandafter \@secondoftwo
 \fi
}%
\providecommand \@ifx [1]{%
 \ifx #1\expandafter \@firstoftwo
 \else \expandafter \@secondoftwo
 \fi
}%
\providecommand \natexlab [1]{#1}%
\providecommand \enquote  [1]{``#1''}%
\providecommand \bibnamefont  [1]{#1}%
\providecommand \bibfnamefont [1]{#1}%
\providecommand \citenamefont [1]{#1}%
\providecommand \href@noop [0]{\@secondoftwo}%
\providecommand \href [0]{\begingroup \@sanitize@url \@href}%
\providecommand \@href[1]{\@@startlink{#1}\@@href}%
\providecommand \@@href[1]{\endgroup#1\@@endlink}%
\providecommand \@sanitize@url [0]{\catcode `\\12\catcode `\$12\catcode
  `\&12\catcode `\#12\catcode `\^12\catcode `\_12\catcode `\%12\relax}%
\providecommand \@@startlink[1]{}%
\providecommand \@@endlink[0]{}%
\providecommand \url  [0]{\begingroup\@sanitize@url \@url }%
\providecommand \@url [1]{\endgroup\@href {#1}{\urlprefix }}%
\providecommand \urlprefix  [0]{URL }%
\providecommand \Eprint [0]{\href }%
\providecommand \doibase [0]{http://dx.doi.org/}%
\providecommand \selectlanguage [0]{\@gobble}%
\providecommand \bibinfo  [0]{\@secondoftwo}%
\providecommand \bibfield  [0]{\@secondoftwo}%
\providecommand \translation [1]{[#1]}%
\providecommand \BibitemOpen [0]{}%
\providecommand \bibitemStop [0]{}%
\providecommand \bibitemNoStop [0]{.\EOS\space}%
\providecommand \EOS [0]{\spacefactor3000\relax}%
\providecommand \BibitemShut  [1]{\csname bibitem#1\endcsname}%
\let\auto@bib@innerbib\@empty
\bibitem [{\citenamefont {Feynman}(1982)}]{feynman1982simulating}%
  \BibitemOpen
  \bibfield  {author} {\bibinfo {author} {\bibfnamefont {R.}~\bibnamefont
  {Feynman}},\ }\href@noop {} {\bibfield  {journal} {\bibinfo  {journal}
  {International journal of theoretical physics}\ }\textbf {\bibinfo {volume}
  {21}},\ \bibinfo {pages} {467} (\bibinfo {year} {1982})}\BibitemShut
  {NoStop}%
\bibitem [{\citenamefont {Hey}(1998)}]{hey1998feynman}%
  \BibitemOpen
  \bibinfo {editor} {\bibfnamefont {A.~J.}\ \bibnamefont {Hey}},\ ed.,\
  \href@noop {} {\emph {\bibinfo {title} {Feynman and Computation}}}\ (\bibinfo
   {publisher} {Perseus Books},\ \bibinfo {address} {Reading},\ \bibinfo {year}
  {1998})\BibitemShut {NoStop}%
\bibitem [{\citenamefont {Stachel}(1986)}]{stachel1986einstein}%
  \BibitemOpen
  \bibfield  {author} {\bibinfo {author} {\bibfnamefont {J.}~\bibnamefont
  {Stachel}},\ }in\ \href@noop {} {\emph {\bibinfo {booktitle} {From Quarks to
  Quasars: Philosophical Problems of Modern Physics}}},\ \bibinfo {editor}
  {edited by\ \bibinfo {editor} {\bibfnamefont {R.~G.}\ \bibnamefont
  {Colodny}}\ and\ \bibinfo {editor} {\bibfnamefont {A.}~\bibnamefont
  {Coffa}}}\ (\bibinfo  {publisher} {University of Pittsburgh Press},\ \bibinfo
  {address} {Pittsburgh},\ \bibinfo {year} {1986})\ pp.\ \bibinfo {pages}
  {349--81}\BibitemShut {NoStop}%
\bibitem [{\citenamefont {Immirzi}(1997)}]{immirzi1997quantum}%
  \BibitemOpen
  \bibfield  {author} {\bibinfo {author} {\bibfnamefont {G.}~\bibnamefont
  {Immirzi}},\ }\href@noop {} {\bibfield  {journal} {\bibinfo  {journal}
  {Nuclear Physics B-Proceedings Supplements}\ }\textbf {\bibinfo {volume}
  {57}},\ \bibinfo {pages} {65} (\bibinfo {year} {1997})}\BibitemShut {NoStop}%
\bibitem [{\citenamefont {Perez}(2003)}]{perez2003spin}%
  \BibitemOpen
  \bibfield  {author} {\bibinfo {author} {\bibfnamefont {A.}~\bibnamefont
  {Perez}},\ }\href@noop {} {\bibfield  {journal} {\bibinfo  {journal}
  {Classical and Quantum Gravity}\ }\textbf {\bibinfo {volume} {20}},\ \bibinfo
  {pages} {R43} (\bibinfo {year} {2003})}\BibitemShut {NoStop}%
\bibitem [{\citenamefont {Bombelli}\ \emph {et~al.}(1987)\citenamefont
  {Bombelli}, \citenamefont {Lee}, \citenamefont {Meyer},\ and\ \citenamefont
  {Sorkin}}]{bombelli1987space}%
  \BibitemOpen
  \bibfield  {author} {\bibinfo {author} {\bibfnamefont {L.}~\bibnamefont
  {Bombelli}}, \bibinfo {author} {\bibfnamefont {J.}~\bibnamefont {Lee}},
  \bibinfo {author} {\bibfnamefont {D.}~\bibnamefont {Meyer}}, \ and\ \bibinfo
  {author} {\bibfnamefont {R.}~\bibnamefont {Sorkin}},\ }\href@noop {}
  {\bibfield  {journal} {\bibinfo  {journal} {Physical review letters}\
  }\textbf {\bibinfo {volume} {59}},\ \bibinfo {pages} {521} (\bibinfo {year}
  {1987})}\BibitemShut {NoStop}%
\bibitem [{\citenamefont {Schild}(1947)}]{Schild}%
  \BibitemOpen
  \bibfield  {author} {\bibinfo {author} {\bibfnamefont {A.}~\bibnamefont
  {Schild}},\ }\href@noop {} {\bibfield  {journal} {\bibinfo  {journal} {Phys.
  Rev.}\ }\textbf {\bibinfo {volume} {74}} (\bibinfo {year}
  {1947})}\BibitemShut {NoStop}%
\bibitem [{\citenamefont {Amelino-Camelia}(2001)}]{amelino2001testable}%
  \BibitemOpen
  \bibfield  {author} {\bibinfo {author} {\bibfnamefont {G.}~\bibnamefont
  {Amelino-Camelia}},\ }\href@noop {} {\bibfield  {journal} {\bibinfo
  {journal} {Physics Letters B}\ }\textbf {\bibinfo {volume} {510}},\ \bibinfo
  {pages} {255} (\bibinfo {year} {2001})}\BibitemShut {NoStop}%
\bibitem [{\citenamefont {Amelino-Camelia}(2002)}]{amelino2002relativity}%
  \BibitemOpen
  \bibfield  {author} {\bibinfo {author} {\bibfnamefont {G.}~\bibnamefont
  {Amelino-Camelia}},\ }\href@noop {} {\bibfield  {journal} {\bibinfo
  {journal} {International Journal of Modern Physics D}\ }\textbf {\bibinfo
  {volume} {11}},\ \bibinfo {pages} {35} (\bibinfo {year} {2002})}\BibitemShut
  {NoStop}%
\bibitem [{\citenamefont {Magueijo}\ and\ \citenamefont
  {Smolin}(2002)}]{magueijo2002lorentz}%
  \BibitemOpen
  \bibfield  {author} {\bibinfo {author} {\bibfnamefont {J.}~\bibnamefont
  {Magueijo}}\ and\ \bibinfo {author} {\bibfnamefont {L.}~\bibnamefont
  {Smolin}},\ }\href@noop {} {\bibfield  {journal} {\bibinfo  {journal}
  {Physical Review Letters}\ }\textbf {\bibinfo {volume} {88}},\ \bibinfo
  {pages} {190403} (\bibinfo {year} {2002})}\BibitemShut {NoStop}%
\bibitem [{\citenamefont {Schumacher}\ and\ \citenamefont
  {Werner}(2004)}]{schumacher2004reversible}%
  \BibitemOpen
  \bibfield  {author} {\bibinfo {author} {\bibfnamefont {B.}~\bibnamefont
  {Schumacher}}\ and\ \bibinfo {author} {\bibfnamefont {R.}~\bibnamefont
  {Werner}},\ }\href@noop {} {\bibfield  {journal} {\bibinfo  {journal} {Arxiv
  preprint quant-ph/0405174}\ } (\bibinfo {year} {2004})}\BibitemShut {NoStop}%
\bibitem [{\citenamefont {Arrighi}\ \emph {et~al.}(2011)\citenamefont
  {Arrighi}, \citenamefont {Nesme},\ and\ \citenamefont
  {Werner}}]{arrighi2011unitarity}%
  \BibitemOpen
  \bibfield  {author} {\bibinfo {author} {\bibfnamefont {P.}~\bibnamefont
  {Arrighi}}, \bibinfo {author} {\bibfnamefont {V.}~\bibnamefont {Nesme}}, \
  and\ \bibinfo {author} {\bibfnamefont {R.}~\bibnamefont {Werner}},\
  }\href@noop {} {\bibfield  {journal} {\bibinfo  {journal} {Journal of
  Computer and System Sciences}\ }\textbf {\bibinfo {volume} {77}},\ \bibinfo
  {pages} {372} (\bibinfo {year} {2011})}\BibitemShut {NoStop}%
\bibitem [{\citenamefont {Meyer}(1996)}]{meyer1996quantum}%
  \BibitemOpen
  \bibfield  {author} {\bibinfo {author} {\bibfnamefont {D.}~\bibnamefont
  {Meyer}},\ }\href@noop {} {\bibfield  {journal} {\bibinfo  {journal} {Journal
  of Statistical Physics}\ }\textbf {\bibinfo {volume} {85}},\ \bibinfo {pages}
  {551} (\bibinfo {year} {1996})}\BibitemShut {NoStop}%
\bibitem [{\citenamefont {Ambainis}\ \emph {et~al.}(2001)\citenamefont
  {Ambainis}, \citenamefont {Bach}, \citenamefont {Nayak}, \citenamefont
  {Vishwanath},\ and\ \citenamefont {Watrous}}]{ambainis2001one}%
  \BibitemOpen
  \bibfield  {author} {\bibinfo {author} {\bibfnamefont {A.}~\bibnamefont
  {Ambainis}}, \bibinfo {author} {\bibfnamefont {E.}~\bibnamefont {Bach}},
  \bibinfo {author} {\bibfnamefont {A.}~\bibnamefont {Nayak}}, \bibinfo
  {author} {\bibfnamefont {A.}~\bibnamefont {Vishwanath}}, \ and\ \bibinfo
  {author} {\bibfnamefont {J.}~\bibnamefont {Watrous}},\ }in\ \href@noop {}
  {\emph {\bibinfo {booktitle} {Proceedings of the thirty-third annual ACM
  symposium on Theory of computing}}}\ (\bibinfo {organization} {ACM},\
  \bibinfo {year} {2001})\ pp.\ \bibinfo {pages} {37--49}\BibitemShut {NoStop}%
\bibitem [{\citenamefont {D'Ariano}\ and\ \citenamefont
  {Perinotti}(2014)}]{PhysRevA.90.062106}%
  \BibitemOpen
  \bibfield  {author} {\bibinfo {author} {\bibfnamefont {G.~M.}\ \bibnamefont
  {D'Ariano}}\ and\ \bibinfo {author} {\bibfnamefont {P.}~\bibnamefont
  {Perinotti}},\ }\href@noop {} {\bibfield  {journal} {\bibinfo  {journal}
  {Phys. Rev. A}\ }\textbf {\bibinfo {volume} {90}},\ \bibinfo {pages} {062106}
  (\bibinfo {year} {2014})}\BibitemShut {NoStop}%
\bibitem [{\citenamefont {Susskind}(1977)}]{PhysRevD.16.3031}%
  \BibitemOpen
  \bibfield  {author} {\bibinfo {author} {\bibfnamefont {L.}~\bibnamefont
  {Susskind}},\ }\href@noop {} {\bibfield  {journal} {\bibinfo  {journal}
  {Phys. Rev. D}\ }\textbf {\bibinfo {volume} {16}},\ \bibinfo {pages} {3031}
  (\bibinfo {year} {1977})}\BibitemShut {NoStop}%
\bibitem [{\citenamefont {Amelino-Camelia}\ and\ \citenamefont
  {Piran}(2001)}]{amelino2001planck}%
  \BibitemOpen
  \bibfield  {author} {\bibinfo {author} {\bibfnamefont {G.}~\bibnamefont
  {Amelino-Camelia}}\ and\ \bibinfo {author} {\bibfnamefont {T.}~\bibnamefont
  {Piran}},\ }\href@noop {} {\bibfield  {journal} {\bibinfo  {journal}
  {Physical Review D}\ }\textbf {\bibinfo {volume} {64}},\ \bibinfo {pages}
  {036005} (\bibinfo {year} {2001})}\BibitemShut {NoStop}%
\bibitem [{\citenamefont {D'Ariano}(2012)}]{mauro2012quantum}%
  \BibitemOpen
  \bibfield  {author} {\bibinfo {author} {\bibfnamefont {G.~M.}\ \bibnamefont
  {D'Ariano}},\ }\href@noop {} {\bibfield  {journal} {\bibinfo  {journal}
  {Physics Letters. Section A: General, Atomic and Solid State Physics}\
  }\textbf {\bibinfo {volume} {376}},\ \bibinfo {pages} {697} (\bibinfo {year}
  {2012})}\BibitemShut {NoStop}%
\bibitem [{\citenamefont {Hawking}(1996)}]{PhysRevD.53.3099}%
  \BibitemOpen
  \bibfield  {author} {\bibinfo {author} {\bibfnamefont {S.~W.}\ \bibnamefont
  {Hawking}},\ }\href@noop {} {\bibfield  {journal} {\bibinfo  {journal} {Phys.
  Rev. D}\ }\textbf {\bibinfo {volume} {53}},\ \bibinfo {pages} {3099}
  (\bibinfo {year} {1996})}\BibitemShut {NoStop}%
\bibitem [{\citenamefont {Moyer}(2012)}]{Moyer:2012ws}%
  \BibitemOpen
  \bibfield  {author} {\bibinfo {author} {\bibfnamefont {M.}~\bibnamefont
  {Moyer}},\ }\href@noop {} {\bibfield  {journal} {\bibinfo  {journal}
  {Scientific American}\ } (\bibinfo {year} {2012})}\BibitemShut {NoStop}%
\bibitem [{\citenamefont {Hogan}(2012)}]{Hogan:2012ik}%
  \BibitemOpen
  \bibfield  {author} {\bibinfo {author} {\bibfnamefont {C.}~\bibnamefont
  {Hogan}},\ }\href@noop {} {\bibfield  {journal} {\bibinfo  {journal}
  {Physical Review D}\ }\textbf {\bibinfo {volume} {85}} (\bibinfo {year}
  {2012})}\BibitemShut {NoStop}%
\bibitem [{\citenamefont {Pikovski}\ \emph {et~al.}(2012)\citenamefont
  {Pikovski}, \citenamefont {Vanner}, \citenamefont {Aspelmeyer}, \citenamefont
  {Kim},\ and\ \citenamefont {Brukner}}]{pikovski2011probing}%
  \BibitemOpen
  \bibfield  {author} {\bibinfo {author} {\bibfnamefont {I.}~\bibnamefont
  {Pikovski}}, \bibinfo {author} {\bibfnamefont {M.}~\bibnamefont {Vanner}},
  \bibinfo {author} {\bibfnamefont {M.}~\bibnamefont {Aspelmeyer}}, \bibinfo
  {author} {\bibfnamefont {M.}~\bibnamefont {Kim}}, \ and\ \bibinfo {author}
  {\bibfnamefont {C.}~\bibnamefont {Brukner}},\ }\href@noop {} {\bibfield
  {journal} {\bibinfo  {journal} {Nature Physics}\ }\textbf {\bibinfo {volume}
  {331}},\ \bibinfo {pages} {393} (\bibinfo {year} {2012})}\BibitemShut
  {NoStop}%
\bibitem [{\citenamefont {Amelino-Camelia}\ \emph {et~al.}(1998)\citenamefont
  {Amelino-Camelia}, \citenamefont {Ellis}, \citenamefont {Mavromatos},
  \citenamefont {Nanopoulos},\ and\ \citenamefont
  {Sarkar}}]{Amelino-Camelia:1998aa}%
  \BibitemOpen
  \bibfield  {author} {\bibinfo {author} {\bibfnamefont {G.}~\bibnamefont
  {Amelino-Camelia}}, \bibinfo {author} {\bibfnamefont {J.}~\bibnamefont
  {Ellis}}, \bibinfo {author} {\bibfnamefont {N.~E.}\ \bibnamefont
  {Mavromatos}}, \bibinfo {author} {\bibfnamefont {D.~V.}\ \bibnamefont
  {Nanopoulos}}, \ and\ \bibinfo {author} {\bibfnamefont {S.}~\bibnamefont
  {Sarkar}},\ }\href@noop {} {\bibfield  {journal} {\bibinfo  {journal}
  {Nature}\ }\textbf {\bibinfo {volume} {393}},\ \bibinfo {pages} {763}
  (\bibinfo {year} {1998})}\BibitemShut {NoStop}%
\bibitem [{\citenamefont {Bisio}\ \emph {et~al.}(2016)\citenamefont {Bisio},
  \citenamefont {D'Ariano},\ and\ \citenamefont {Perinotti}}]{Bisio2016}%
  \BibitemOpen
  \bibfield  {author} {\bibinfo {author} {\bibfnamefont {A.}~\bibnamefont
  {Bisio}}, \bibinfo {author} {\bibfnamefont {G.~M.}\ \bibnamefont {D'Ariano}},
  \ and\ \bibinfo {author} {\bibfnamefont {P.}~\bibnamefont {Perinotti}},\
  }\href@noop {} {\bibfield  {journal} {\bibinfo  {journal} {Annals of
  Physics}\ ,\ \bibinfo {pages} {http://dx.doi.org/10.1016/j.aop.2016.02.009}}
  (\bibinfo {year} {2016})}\BibitemShut {NoStop}%
\bibitem [{\citenamefont {Vasileiou}\ \emph {et~al.}(2015)\citenamefont
  {Vasileiou}, \citenamefont {Granot}, \citenamefont {Piran},\ and\
  \citenamefont {Amelino-Camelia}}]{Vasileiou:2015}%
  \BibitemOpen
  \bibfield  {author} {\bibinfo {author} {\bibfnamefont {V.}~\bibnamefont
  {Vasileiou}}, \bibinfo {author} {\bibfnamefont {J.}~\bibnamefont {Granot}},
  \bibinfo {author} {\bibfnamefont {T.}~\bibnamefont {Piran}}, \ and\ \bibinfo
  {author} {\bibfnamefont {G.}~\bibnamefont {Amelino-Camelia}},\ }\href@noop {}
  {\bibfield  {journal} {\bibinfo  {journal} {Nat Phys}\ }\textbf {\bibinfo
  {volume} {11}},\ \bibinfo {pages} {344} (\bibinfo {year} {2015})}\BibitemShut
  {NoStop}%
\bibitem [{\citenamefont {Amelino-Camelia}\ \emph {et~al.}(2011)\citenamefont
  {Amelino-Camelia}, \citenamefont {Freidel}, \citenamefont
  {Kowalski-Glikman},\ and\ \citenamefont {Smolin}}]{PhysRevD.84.084010}%
  \BibitemOpen
  \bibfield  {author} {\bibinfo {author} {\bibfnamefont {G.}~\bibnamefont
  {Amelino-Camelia}}, \bibinfo {author} {\bibfnamefont {L.}~\bibnamefont
  {Freidel}}, \bibinfo {author} {\bibfnamefont {J.}~\bibnamefont
  {Kowalski-Glikman}}, \ and\ \bibinfo {author} {\bibfnamefont
  {L.}~\bibnamefont {Smolin}},\ }\href@noop {} {\bibfield  {journal} {\bibinfo
  {journal} {Phys. Rev. D}\ }\textbf {\bibinfo {volume} {84}},\ \bibinfo
  {pages} {084010} (\bibinfo {year} {2011})}\BibitemShut {NoStop}%
\bibitem [{\citenamefont {D'Ariano}\ \emph {et~al.}(2015)\citenamefont
  {D'Ariano}, \citenamefont {Erba}, \citenamefont {Perinotti},\ and\
  \citenamefont {Tosini}}]{d2015virtually}%
  \BibitemOpen
  \bibfield  {author} {\bibinfo {author} {\bibfnamefont {G.~M.}\ \bibnamefont
  {D'Ariano}}, \bibinfo {author} {\bibfnamefont {M.}~\bibnamefont {Erba}},
  \bibinfo {author} {\bibfnamefont {P.}~\bibnamefont {Perinotti}}, \ and\
  \bibinfo {author} {\bibfnamefont {A.}~\bibnamefont {Tosini}},\ }\href@noop {}
  {\bibfield  {journal} {\bibinfo  {journal} {arXiv preprint arXiv:1511.03992}\
  } (\bibinfo {year} {2015})}\BibitemShut {NoStop}%
\end{thebibliography}

\begin{thebibliography}{99}
\bibitem{dirac3d} G. M. D'Ariano, P. Perinotti, Phys. Rev. A {\bf 90}, 062106 (2014).
\bibitem{d2015virtually2}
G. M. D'Ariano,M. Erba, P. Perinotti, A. Tosini, 
 arXiv preprint arXiv:1511.03992.
\bibitem{note1}
Within this context, the isotropy condition, meaning that all the
direction on the lattice are equivalent, is stated as follows.
First we require that the set $S$ is the disjoint union of two subsets $S_+ \cup S_- = S$
such that if $\bvec{h} \in S_+$ then  $\bvec{h} \in S_-$.
Then, let $L$ be a group
automorphisms of the Bravais lattice
which is transitive over the translations in  $S_+$.  We then say that $A$
is isotropic if there exists a faithful representation $U$ of
$L$ over $\mathbb{C}^s$ such that one has
\unexpanded{\begin{align}
  A = \sum_{\bvec{h} \in S } T_{\bvec{h}} \otimes A_{\bvec{h}} = \sum_{\bvec{h} \in S } T_{l(\bvec{h})}
  \otimes  U_l A_{\bvec{h}} U^\dagger_l, \quad \forall l \in L  \nonumber.
\end{align}}

\end{thebibliography}
\end{document}